\documentclass[12pt]{article}
\usepackage{amssymb,amsmath,latexsym}
\title{Delocalization in random polymer models}

\author{S. Jitomirskaya$^1$, H. Schulz-Baldes$^{2}$, G. Stolz$^{3}$
\\
\\
$^1$
{\small Department of Mathematics, University of California at Irvine,
Ca, 92697, USA}
\\
$^2$
{\small Fachbereich Mathematik, Technische Universit{\"a}t Berlin, 
10623, Germany}
\\$^3$
{\small Department of Mathematics,
University of Alabama at Birmingham, Al, 35294 USA}
}

\date{ }

\newtheorem{theo}{Theorem}
\newtheorem{defini}{Definition}
\newtheorem{proposi}{Proposition}
\newtheorem{lemma}{Lemma}

\newtheorem{rem}{Remark}

\newcommand{\CC}{{\mathbb C}}
\newcommand{\NN}{{\mathbb N}}
\newcommand{\RR}{{\mathbb R}}

\newcommand{\ZZ}{{\mathbb Z}}

\newcommand{\PP}{{\bf P}}
\newcommand{\EE}{{\bf E}}

\newcommand{\Ee}{{\cal E}}

\newcommand{\Ss}{{\cal S}}
\newcommand{\Oo}{{\cal O}}

\newcommand{\Nn}{{\cal N}}

\newcommand{\dyn}{{\cal S}_{\epsilon,\omega}}

\newcommand{\dynl}{{\cal S}_{\epsilon,\omega_l}}

\newcommand{\dynpm}{{\cal S}_{\epsilon,\pm}}

\begin{document}

\maketitle

\begin{abstract}
A random polymer model is a one-dimensional Jacobi matrix
randomly composed of two finite building blocks. If the two
associated transfer matrices commute, the corresponding energy is
called critical. Such critical energies appear in physical models, an
example being the widely studied random dimer model. It is proven that
the Lyapunov exponent vanishes quadratically at a generic critical
energy and that the density of states is positive there. Large
deviation estimates around these asymptotics allow to prove optimal
lower bounds on quantum transport, showing that it is almost surely
overdiffusive even though the models are known to have pure-point
spectrum with exponentially localized eigenstates for almost every
configuration of the polymers. Furthermore, the level spacing is
shown to be regular at the critical energy.
\end{abstract}

\vspace{.5cm}

\section{Introduction}

Until quite recently,  common wisdom was that one-dimensional
random Schr{\"o}dinger operators are in a strong localization phase
and that there is nothing else of any interest to be discovered.
In 1990, Dunlap, Wu and Phillips \cite{DWP} studied the random
dimer model. It is similar to the conventional  Bernoulli-Anderson
model with a Hamiltonian given  by the sum of the discrete
Laplacian and a random potential taking only two values, except
that these potential values now always come in neighboring pairs
(dimers).  For suitable values of the parameters, this model has
so-called critical energies at which the two transfer matrices
across the dimers commute.  This leads to a vanishing of the
Lyapunov exponent and a divergence of the localization length at
these energies.  They then argued and showed numerically that  the
second moment of the position operator $X$ on the lattice grows
superdiffusively  under the dynamics like  $\langle \psi
|X^2(t)|\psi\rangle\approx C\,t^{3/2}$ for a localized initial
state $\psi$  and any typical dimer configuration.  This is likely
to be responsible for the high conductivities of certain organic
polymer chains \cite{PW} and quasi-$1D$ semiconductor
superlattices \cite{PTB}.

\vspace{.2cm}

This work considers Jacobi matrices $H_\omega$ ramdomly
composed of two building blocks modeling two different polymers.
The associated polymer transfer matrices are supposed to
commute at an energy lying within the spectra of the
periodic operators containing only one of the polymers.
It is proven that for {\sl almost
every} polymer configuration $\omega$ and every $\alpha>0$ there is a
positive constant $C_\alpha$ such that

\begin{equation}
\label{eq-main}
\int^T_0 \frac{dt}{T}\;
\langle 0 |e^{\imath H t}|X|^q
e^{-\imath H t}|0\rangle
\;\geq\;
C_\alpha\, T^{q-\frac{1}{2}-\alpha}
\mbox{ , }
\end{equation}

\noindent where $|0\rangle$ is the state localized at the origin. 
For $q=2$, this
rigorously confirms that the heuristics of \cite{DWP} (discussed
below) provide a correct lower bound on transport. To prove the
corresponding upper bound for critical energies of order 1 in the
sense of Definition \ref{def-critical2} remains an open problem, 
but we believe the quantitative lower bound on the Lyapunov exponent 
(Theorem \ref{theo-Lyapasym}) to be a central ingredient.
Moreover, it is
shown that for {\sl every} configuration the l.h.s. is greater or equal than
$C\,T^{q-1} $ for some $C>0$. Note that (\ref{eq-main}) implies that
the conductivity is infinite either at finite temperature or if the
critical energy is at the Fermi level \cite{SB}. 

\vspace{.2cm}

The above results should be confronted with the fact that the spectrum of a
random polymer model is almost surely pure-point with
exponentially localized eigenfunctions. For the related
Bernoulli-Anderson model, such {\sl spectral}  localization results
were first proven in \cite{CKM}, later on in \cite{SVW}. More
recently, the random dimer model and the continuous Bernoulli-Anderson
model were treated in \cite{BG} and \cite{DSS1} respectively.
These works also established dynamical localization on energy
intervals not containing a discrete set of special energies which
includes the above critical energies. While \cite{DSS1} considers
continuum models, its approach can be carried over to prove spectral
localization and dynamical localization away from the set of special
energies for the polymer models studied here, see \cite{DSS2} for the
more general case of an arbitrary number of building blocks of bounded length.

\vspace{.2cm}

The fact that spectral
localization can in principle coexist with quantum transport (even
almost ballistic; note that ballistic is impossible with pure point
spectrum \cite{Sim}) was
demonstrated by an example in \cite{RJLS} (see also \cite{BT}).
However, those examples were rather
artificial and much research since then was devoted to the program of
proving dynamical localization (i.e. boundedness in time of the
left-hand side of (\ref{eq-main})) in models of physical interest with
previously established spectral  localization, by means of upgrading
the proof of pure point spectrum (see \cite{GK} and references
therein, also \cite{bj}).  The success of this program may have raised 
doubts as to the
validity of the distinction between spectral and dynamical
localization in physically relevant contexts.  This paper demonstrates
that the distinction should indeed be made as it shows  that
exponential localization and quantum transport coexist also in
physical models.

\vspace{.2cm}

Let us sketch the heuristics of \cite{DWP} leading to (\ref{eq-main}). 
It is known \cite{Bov} and proven below that the Lyapunov exponent
generically vanishes quadratically like
$\gamma(E_c+\epsilon)=c\epsilon^2+\Oo(\epsilon^3)$ in the vicinity
of the critical energy $E_c$. The extension of the eigenstates in
an $\epsilon$-neighborhood of $E_c$ is given by their localization
length equal to the inverse of the Lyapunov exponent. Therefore
the portion of the initial wave packet lying energetically in this
$\epsilon$-neighborhood spreads out ballistically up to time
scales $T\approx \epsilon^{-2}$. Because it will be shown that the
density of states is positive at $E_c$, this portion of states is
proportional to $\epsilon$. Consequently, the $q$th moment of the
position operator should grow like $T^q$ multiplied by this factor
$\epsilon\approx T^{-1/2}$, showing that (\ref{eq-main}) should hold
with high probability.

\vspace{.2cm}

The main technical tools are adequate action-angle variables, also
called modified Pr{\"u}fer variables in the mathematical
literature. Adapting techniques from \cite{PF}, one obtains
perturbative expansions around the critical energy of both the density
of states and the Lyapunov exponent, which prove positivity of the
density of states and  quadratic vanishing of the Lyapunov exponent
near a generic critical energy. The proof of  (\ref{eq-main}) requires
an additional large deviations analysis for the Lyapunov exponent. The
methods of proof are calculatory,  quantitative and optimal.  For
example, they allow to show how large the moments of the position
operator have to be if the commutator of the transfer matrices is
small, but does not vanish.

\vspace{.2cm}

\noindent {\bf Acknowledgements:} This paper is a heavily revised
version of a preprint \cite{JSS} which
contained a first, but less direct proof of the deterministic lower
bound stated in Theorem \ref{theo-lowerdeter} below. The basic
strategy (Lemma \ref{lem-transferperturb} and Section \ref{sec-dynamics})
of the present proof of the dynamical lower
bound (Theorem \ref{theo-lower}) given the boundedness of transfer
matrices (Theorem \ref{btt}) was suggested by S. 
Tcheremchantsev. This technique, 
which will
be published in full generality in \cite{DT}, is simpler than the Guarneri
method \cite{Gua} of proving lower bounds employed in \cite{JSS} 
(and also applicable here) and allowed us to circumvent
previous more intricate arguments. We greatly appreciate that
S. Tcheremchantsev made his work available prior to publication.
S. J. and H. S.-B. were supported by NSF grant DMS-0070755, H.
S.-B. moreover by DFG grant SCHU 1358/1-1 and the SFB 288. G. S. was
supported by NSF grant DMS-0070343. He would also like to acknowledge
financial support of CNRS (France) and hospitality at Universit{\'e}
Paris 7, where part of this work was done. 

\vspace{.2cm}

\section{Model and main results}
\label{sec-res}

Let $\hat{t}_\pm=(\hat{t}_\pm(0),\ldots,\hat{t}_\pm(L_\pm-1))$ and
$\hat{v}_\pm=(\hat{v}_\pm(0),\ldots,\hat{v}_\pm(L_\pm-1))$ be two
pairs of finite sequences of real numbers, satisfying
$\hat{t}_\pm(l)>0$ for all $l=0,\ldots, L_\pm-1$. These numbers are
the hopping and potential terms of two different polymers. A family
of random Jacobi matrices is now constructed by random juxtaposition
of these polymers. More precisely,
to any sequence $\omega=(\omega
_l)_{l\in\ZZ}$ of signs $+$ and $-$ one associates
sequences
$t_\omega=(t_\omega(n))_{n\in\ZZ}$ and
$v_\omega=(v_\omega(n))_{n\in\ZZ}$ by means of $t_\omega=(\ldots,
\hat{t}_{\omega_0}, \hat{t}_{\omega_1},\ldots)$ and
$v_\omega=(\ldots, \hat{v}_{\omega_0}, \hat{v}_{\omega_1},\ldots)$.
An exact definition of the underlying probability space $(\Omega,
\PP)$, which also requires to randomize the position of
$\hat{v}_{\omega_0}(0)$ and $\hat{t}_{\omega_0}(0)$, is given in 
Section~\ref{sec-polymers}.
The polymer
Hamiltonian $H_\omega$ of the configuration $\omega$
is then defined by

\begin{equation}
\label{eq-polymerHam}
(H_\omega\psi)(n)
\;=\;
-t_\omega(n+1) \psi(n+1)+v_\omega(n)\psi(n)-t_\omega(n)\psi(n-1)
\mbox{ , }
\qquad
\psi\in\ell^2(\ZZ)
\mbox{ , }
\end{equation}

\noindent  and $(H_\omega)_{\omega\in\Omega}$
becomes a family of random operators if the signs are chosen with
probabilities $p_+$ and $p_-=1-p_+$ respectively.
The polymer transfer matrices $T^E_\pm$ at energy
$E\in\RR$ are introduced by

\begin{equation}
\label{eq-transfer}
T^E_\pm
\;=\;
T_{\hat{v}_\pm(L_{\pm}-1)-E,\hat{t}_\pm(L_{\pm}-1)} \ldots 
T_{\hat{v}_\pm(0)-E,\hat{t}_\pm(0)}
\mbox{ , }
\qquad \mbox{where }\;\;\;\;\;
T_{v,t}\;=\;\frac{1}{t}\,
\left(\begin{array}{cc} v & -t^2 \\ 1 & 0 \end{array} \right)
\mbox{ . }
\end{equation}

\noindent The transfer matrices over several polymers are then

\begin{equation}
\label{eq-transsevpol}
T_{\omega}^E(k,m) = T_{\omega_{k-1}}^E \cdot T_{\omega_{k-2}}^E \cdot
\ldots \cdot T_{\omega_m}^E
\mbox{ , }
\qquad k>m 
\mbox{ , } 
\end{equation}

\noindent and $T_{\omega}^E(k,m) = T_{\omega}^E(m,k)^{-1}$ if $k<m$,
$T_{\omega}^E(m,m)={\bf 1}$. The Lyapunov exponent at energy $E$,
also called inverse localization
length, is then almost surely
defined by (some more details are given in Sections \ref{sec-covJac}
and \ref{sec-polymers})

\begin{equation}
\label{eq-Lyapdef}
\gamma(E)
\;=\;
\lim_{k\to\infty}\frac{1}{k\,\langle L_\pm\rangle}
\log\left(\left\|T^E_{\omega}(k,0)
\right\|\right)\mbox{ , }
\end{equation}

\noindent where $\langle c_\pm\rangle=p_+c_++p_-c_-$ for any complex
numbers $c_\pm$.
Vanishing of the Lyapunov exponent is considered an indicator for
possible delocalization. For a polymer chain, this happens in the following
situation:

\begin{defini}
\label{def-critical}
An energy $E_c\in\RR$ is called critical for the random family
$(H_\omega)_{\omega\in\Omega}$ of polymer Hamiltonians if
the polymer transfer matrices  $T_\pm^{E_c}$ are elliptic {\rm (}i.e.
$|\mbox{\rm Tr}(T_\pm^{E_c})|<2${\rm )} or equal to $\pm{\bf 1}$ and
commute
\begin{equation}
\label{eq-critical}
[T^{E_c}_-,T^{E_c}_+]\;=\;0\mbox{ . }
\end{equation}
\end{defini}

\begin{rem}
{\rm
The definition does {\sl not} allow the critical energy to be in a
spectral gap or at the
band edges of one of the periodic operators (constructed from $(\hat{t}_+
,\hat{v}_+)$  and $(\hat{t}_-,\hat{v}_-)$ respectively) except for
points of band touching (where the transfer matrix is $\pm{\bf 1}$).
}
\end{rem}

\begin{rem}
{\rm
The condition (\ref{eq-critical}) contains 4 equations.
Given a model, one can only vary the energy. Hence, in the space of
polymer models existence of critical energies is a non-generic
property. On the other hand, given an energy $E_c$,
it is always possible to construct polymer models
that have $E_c$ as a critical energy. 
}
\end{rem}

\vspace{.2cm}

\noindent {\bf Examples} If $L_\pm=1$, the model reduces to the
Bernoulli-Anderson model and there are no critical
energies. If $L_+=2$ and $L_-=1$, an example can be constructed as
follows: choose $t(l)=1$ for all $l\in\ZZ$ and $\hat{v}_+=(0,0)$ and
$\hat{v}_-=(\lambda)$ and $|\lambda|<2$, then $E_c=0$
is the critical energy.
The most prominent \cite{DWP,Bov,BG} example is
the random dimer model for which $L_+=L_-=2$ and $\hat{v}_+(0)=
\hat{v}_+(1)=\lambda$ and
$\hat{v}_-(0)=\hat{v}_-(1)=-\lambda$ ($\lambda\in\RR$),
and $t(l)=1$ for all $l\in\ZZ$.
This model has two critical energies $E_c=\lambda$
and $E_c=-\lambda$ as long as $\lambda<1$.
It was previously (non-rigorously) known that
$\gamma(E_c+\epsilon)=\Oo(\epsilon^2)$ \cite{Bov} for the random dimer
model.

\vspace{.2cm}

The definition of the critical energy
assures that there exists a real invertible
matrix $M$ transforming $T^{E_c}_-$ and  $T^{E_c}_+$
simultaneously into rotations by angles $\eta_-$ and
$\eta_+$ respectively:

\begin{equation}
\label{eq-tric}
MT^{E_c}_{\pm}M^{-1}
\;=\;
\left(\begin{array}{cc}
\cos(\eta_\pm) & -\sin(\eta_\pm)
\\ \sin(\eta_\pm)
& \cos(\eta_\pm)
\end{array} \right)
\mbox{ . }
\end{equation}

\noindent
Hence $\gamma(E_c)=0$. Because $T_{\pm}^E$ are polynomials of
degree $L_{\pm}$ in $E$, one can expand $T_{\pm}^{E_c+\epsilon}$
into powers of $\epsilon$. Since $MT_{\pm}^{E_c} M^{-1}$ are
rotations, this implies that $\|MT_{\pm}^{E_c+\epsilon} M^{-1}\|
\leq 1 + c|\epsilon|$ for $|\epsilon| \le \epsilon_0$ and one deduces
the following:

\begin{proposi}
\label{prop-Lyaptrivial}
For $\epsilon_0 >0$ there exists a constant $C<\infty$ such
that for all $|\epsilon| \le \epsilon_0$ and $m,k\in \ZZ$,
\begin{equation}
\label{eq-lyapbound}
\left\|
T^{E_c+\epsilon}_{\omega}(k,m)\right\|
\;\le \;
C \,e^{C|\epsilon|\,|k-m|}
\mbox{ . }
\end{equation}
\noindent In particular, 
$|\gamma(E_c+\epsilon)|\leq \,C'\,|\epsilon|$ for $C'>0$.
\end{proposi}

Note that the bound in (\ref{eq-lyapbound}) does not depend on the
configuration. 
To study a possible spreading of wave packets due to the divergence of
the localization length, one best
considers the moments of the associated probability distribution,
notably the time-averaged 
moments of the position operator $X$ on $\ell^2(\ZZ)$:

\begin{equation}
\label{eq-moment}
M_{\omega,q}(T)
\;=\;
\int^\infty_{0}\frac{dt}{T}\,e^{-\frac{t}{T}}
\;\langle 0 |
e^{\imath H_\omega t}|X|^q e^{-\imath H_\omega t}
|0\rangle
\mbox{ , }
\qquad
q>0
\mbox{ , }
\end{equation}

\noindent The exponential time average may be replaced by a Cesaro
mean without changing the asymptotics ({\sl e.g.} \cite{GSB}). 
Proposition \ref{prop-Lyaptrivial} will lead more or less
directly to the following deterministic lower bound on transport.

\begin{theo}
\label{theo-lowerdeter}
There exists a constant $C$ such that for every configuration $\omega$
and for $q\geq 0$

\begin{equation}
\label{eq-lowerdeter}
M_{\omega,q}(T)
\;\geq\;
C\,T^{q-1}
\mbox{ . }
\end{equation}

\end{theo}

\begin{rem}
{\rm
It is important that the initial condition in (\ref{eq-moment})
is $|0\rangle$ and not an arbitrary state $\psi \in \ell^2(\ZZ)$. In
fact, $\psi$ could be an eigenstate of $H_\omega$ and hence not lead
to any diffusion.
}
\end{rem}

\vspace{.2cm}

In order to study the behavior of the Lyapunov
exponent in the vicinity of the critical energy, that is, go beyond
the trivial upper bound $|\gamma(E_c+\epsilon)|\leq \,C'\,|\epsilon|$, 
let us define the
transmission and reflection coefficients $a^\epsilon_\pm$ and
$b^\epsilon_\pm$ by

\begin{equation}
\label{eq-reflec}
MT^{E_c+\epsilon}_{\pm}M^{-1}\,v
\;=\;
a^\epsilon_\pm v + b^\epsilon_\pm \overline{v}
\mbox{ , }
\qquad
v\;=\;
\frac{1}{\sqrt{2}}
\left(\begin{array}{c}
1 \\ -\imath
\end{array} \right)
\mbox{ . }
\end{equation}

\noindent Both are polynomials in $\epsilon$. 
As $v$ is an eigenvector
of all rotations, one has

\begin{equation}
\label{ab0}
a_\pm^0\;=\;e^{\imath \eta_\pm}\mbox{ , }
\qquad
b^0_\pm\;=\;0
\mbox{ . }
\end{equation}

\noindent Furthermore let us set 
$e^{\imath\eta^\epsilon_\pm}=a^\epsilon_\pm/|a^\epsilon_\pm|$ so that
$\eta^0_\pm=\eta_\pm$. Averages will always be denoted as
$\langle c_\pm\rangle=p_+c_++p_-c_-$.

\begin{theo}
\label{theo-Lyapasym}
Suppose that $\langle e^{2\imath\eta_\pm}\rangle\neq
1$ and $\langle e^{4\imath\eta_\pm}\rangle\neq
1$. Then the Lyapunov exponent of a random polymer chain satisfies
\begin{equation}
\label{eq-lyapres} \gamma(E_c+\epsilon) \; =\;
\frac{2\,p_+p_-}{\langle L_\pm\rangle}\;
\frac{|b_+^{\epsilon}\,\sin(\eta^\epsilon_-)-
b_-^{\epsilon}\,\sin(\eta^\epsilon_+)|^2}{
|1-\langle e^{2\imath \eta^\epsilon_\pm}\rangle|^2}
+\Oo\left(|b^\epsilon_\pm|^3\right) \mbox{ . }
\end{equation}

\end{theo}

\vspace{.2cm}

\begin{defini}
\label{def-critical2}
A critical energy $E_c\in\RR$  of a polymer Hamiltonian is said to be
of order $r$ if $|b^\epsilon_\pm|=\Oo(\epsilon^r)$,
but not $|b^\epsilon_\pm|=\Oo(\epsilon^{r+1})$ for both polymers.
\end{defini}

\begin{rem}
{\rm If $E_c$ is a critical energy of order $r$, then
$\gamma(E_c+\epsilon)=C\,\epsilon^{2r}+\Oo(\epsilon^{2r+1})$ for
some non-negative constant $C$. Since $b_{\pm}^{\epsilon} =
\Oo(\epsilon)$, the order of every critical energy is as least
$1$, i.e.\ the Lyapunov exponent vanishes at least quadratically.
Generically the order of a critical energy is $1$ (this is the
case in the dimer model). In the latter case, more explicit
formulas for the coefficient in (\ref{eq-lyapres}) invoking only
the values of $T^{E_c}_\pm$ and $\partial_E T^{E_c}_\pm$ at the critical
energy can easily be written out. 
A comparison of (\ref{eq-lyapres}) with a random phase
approximation is made in Section \ref{subsec-comments}. Finally, let
us note that (\ref{eq-lyapres}) also proves positivity of the Lyapunov
exponent close to $E_c$ whenever the numerator does not vanish.
}
\end{rem}

\begin{rem}
{\rm 
The conditions $\langle e^{2\imath\eta_\pm}\rangle\neq 1$ and
$\langle e^{4\imath\eta_\pm}\rangle\neq 1$ in Theorems
\ref{theo-Lyapasym} and \ref{theo-DOSasym} below are linked to
anomalies studied in \cite{CK}.
For the dimer model, the condition 
$\langle e^{4\imath\eta_\pm}\rangle\neq 1$ is verified if and only if
$\lambda\neq 1/\sqrt{2}$. This particular value already appeared in
\cite{BG}. 
}
\end{rem}

Theorem \ref{theo-Lyapasym} shows how the localization length diverges
at the critical energy. The next result concerns the asymptotics of the
integrated density of states $\Nn$ (denoted IDS, its definition is
recalled in Section \ref{sec-covJac} below).
Let $\Nn_\pm$ and $\Nn_\pm'$ denote the absolutely continuous IDS and
their densities
associated to the models with $L_\pm$-periodic models composed of only
one of the polymers. By definition of a critical energy, 
$\Nn_\pm'(E_c)>0$.

\begin{theo}
\label{theo-DOSasym}
Suppose that $\langle e^{2\imath\eta_\pm}\rangle\neq
1$. Then the {\rm IDS} of a random polymer chain satisfies

\begin{equation}
\label{eq-IDSexpan}
\Nn(E_c+\epsilon)
\;=\;
\frac{\langle L_\pm\,\Nn_\pm(E_c)\rangle}{\langle L_\pm\rangle}
+
\epsilon\;
\frac{\langle L_\pm\,\Nn_\pm'(E_c)\rangle}{\langle L_\pm\rangle}
+\Oo(\epsilon^2)
\mbox{ . }
\end{equation}

\end{theo}

\vspace{.2cm}

The theorem states that $\Nn$ is  linearly increasing at
$E_c$ so that there are many states in the vicinity of a critical
energy. The spreading of these states is quantitatively nicely
characterized by the diffusion exponents, namely the power law
growth exponents of the moments $M_{\omega,q}(T)$ defined in
(\ref{eq-moment}) above:

\begin{equation}
\label{eq-exponent}
\beta^\pm_{\omega,q}
\;=\;
\lim_{T\to\infty}\,^{\!\!\!\!\pm}
\frac{\log(M_{\omega,q}(T))}{\log(T^q)}
\mbox{ . }
\end{equation}

\noindent Here $\lim^{\pm}$ denote the superior and inferior limit
respectively. The main result is the following:

\begin{theo}
\label{theo-lower}
Suppose that $|\langle e^{2\imath\eta_\pm}\rangle|<
1$. Then $\PP$-almost surely

\begin{equation}
\label{eq-lower}
\beta^\pm_{\omega,q}
\;\geq\;
1-\frac{1}{2q}
\mbox{ . }
\end{equation}

\end{theo}

\vspace{.2cm}

It is interesting to compare Theorems \ref{theo-lower} and
\ref{theo-lowerdeter}. The latter implies for all configurations
a weaker
lower bound in (\ref{eq-lower}) of the form $1-\frac{1}{q}.$ One can
construct configurations $\omega$ with slower transport than in
(\ref{eq-lower}). Therefore - seemingly paradoxically - 
typical random configurations
do not lead the slowest possible transport for this model.

\vspace{.2cm}

Finally it is worth mentioning a large deviation result here.
The IDS and Lyapunov exponent are both averaged quantities describing
the behavior at the infinite volume limit. Given their asymptotics
$\Nn(E_c+\epsilon)=\Nn(E_c)+\Nn'(E_c)\epsilon+\Oo(\epsilon^2)$
and $\gamma(E_c+\epsilon)=C\epsilon^{2}+\Oo(\epsilon^{3})$ (here $C=0$
if the order of $E_c$ is $p>1$), one
therefore expects that typically (w.r.t. $\PP$) the
following holds for the
finite (but sufficiently large) size Hamiltonian $H_{\omega,N}$ found
by restricting $H_{\omega}$ to $\ell^2(\{0,\ldots,N-1\})$ (with
Dirichlet boundary conditions):
$H_{\omega,N}$ has $cN^{1/2}$
equally spaced eigenstates in the interval
$[E_c-N^{-1/2},E_c+N^{-1/2}]$ which are all spread out over the whole sample.
Here we give an upper bound on the probability of the set of atypical
configurations for which the average {\it metal-like}
behavior of the eigenvalue spacing does not hold.
This result shows on which scales 
there is strong level repulsion.

\begin{theo}
\label{theo-DOSdeviations}
For every $\alpha>0$ there exist $c>0$ and $C<\infty$ such that for
all $N\in\NN$ there are sets
$\Omega_{N}(\alpha)\subset\Omega$ satisfying

$$
\PP(\Omega_{N}(\alpha))
\;=\;
\Oo(e^{-cN^{\alpha}})
\mbox{ , }
$$

\noindent such that for every configuration $\omega$ in the
complementary set $\Omega_{N}(\alpha)^c=\Omega \backslash 
\Omega_{N}(\alpha)$ the following statement holds:
the interval $[E_c-N^{-1/2-\alpha},E_c+N^{-1/2-\alpha}]$
contains of the order of $N^{1/2-\alpha} $ eigenvalues of
$H_{\omega,N}$ which are equally spaced and have eigenfunctions spread
out over the whole sample, namely adjacent eigenvalues $E$ and $E'$ satisfy

\begin{equation}
\label{eq-evspacing}
\frac{1}{C\,N}
\;\leq\;
|E-E'|
\;\leq\;
\frac{C}{N}
\mbox{ , }
\end{equation}

\noindent and for all normalized eigenfunctions $\psi$ of
$H_{\omega,N}$ it holds that

\begin{equation}
\label{eq-efspreading}
\frac{1}{C\,N}
\;\leq\;
|\psi(k-1)|^2 + |\psi(k)|^2 
\;\leq\;
\frac{C}{N}
\end{equation}

\noindent for $0\le k \le N-1$, where $\psi(-1)=\psi(N)=0$.

\end{theo}

Outside of the interval $[E_c-N^{-1/2-\alpha},E_c+N^{-1/2-\alpha}]$ we
expect Poisson statistics. It seems unknown what the level statistics
is like on the boundaries of this interval, but it is possibly not of
the Wigner-type.

\vspace{.2cm}

Theorems \ref{theo-Lyapasym} and \ref{theo-DOSasym} are proved through
the perturbation analysis of polymer phase shifts and action
multipliers (essentially, appropriately modified Pr{\"u}fer
variables).  The key for the proof of Theorems \ref{theo-lower}, and
\ref{theo-DOSdeviations} is Theorem \ref{btt} which states that with
high probability norms of the transfer matrices $T^E_\omega(k,m)$, 
$1\leq m \leq k \leq N$ for energies in  the interval
$[E_c-N^{-1/2-\alpha},E_c+N^{-1/2-\alpha}]$ are uniformly
bounded. This theorem is proved in Section \ref{sec-deviations} by establishing
large-deviation estimates for random Weyl-type sums defined in terms
of polymer phase shifts.

\vspace{.2cm}

Let us conclude with a brief remark about the one-dimensional 
Anderson model in the weak coupling limit, namely
$H_{\lambda,\omega}=H_0+\lambda V_\omega$ where $H_0$ is a periodic
operator, $V_\omega$ the usual Anderson potential and
$\lambda$ a (small) coupling constant.
Pastur and Figotin \cite{PF} showed (in the case where $H_0$ is the
discrete Laplacian) that away from band-center and
band edges of the periodic operator, the Lyapunov exponent grows
quadratically in $\lambda$. The large deviation results and dynamical
lower bounds presented here transpose in order to show that almost
surely 

$$
\sup_{T>0}\,
M_{\omega,\lambda,q}(T)\;\geq\;C_\alpha\,\lambda^{-2q+\alpha}
\mbox{ . }
$$

\noindent Hence the presented techniques allow to study in a very
detailed way the metal-insulator transition
driven by either the disorder strength or the sample size.
This transition appears at a single energy, the critical energy, in
the polymer models studied here.

\vspace{.2cm}

\section{Brief review of basic formulas}
\label{sec-review}

\subsection{Transfer matrices}
\label{sec-transfer}

Let $(t(n))_{n\in\ZZ}$ be a sequence of positive numbers and
$(v(n))_{n\in\ZZ}$ a sequence of real numbers. As in
(\ref{eq-polymerHam}) they define a Jacobi matrix $H$ acting on
$\ell^2(\ZZ)$. Given an initial
angle $\theta^0\in\RR$ and a complex energy $z\in\CC$, let us construct the
formal solution $(u^z(n))_{n\in\ZZ}$ by

\begin{equation}
\label{eq-eigenstates}
-t(n+1)u^z(n+1)+v(n)u^z(n)-t(n)u^z(n-1)
\;=\;
z u^z(n)
\mbox{ , }
\end{equation}

\noindent and the initial conditions

$$
\left(\begin{array}{c} t(0)\,u^z(0) \\ u^z(-1)
\end{array} \right)
\;=\;
\left(\begin{array}{c} \cos(\theta^0) \\ \sin(\theta^0)
\end{array} \right)
\mbox{ . }
$$

\noindent Using the definition (\ref{eq-transfer}) of the single site
transfer matrices $T_{v,t}$, the transfer matrix from site $k$ to $n$
is introduced by

$$
{\cal T}^z(n,k)\;=\;\prod_{l=n-1}^{k}\,T_{v(l)-z,t(l)}
\mbox{ . }
$$

\noindent It allows to rewrite the (formal) eigenfunction equation  
(\ref{eq-eigenstates}) as

\begin{equation}
\label{eq-transferrel}
\left(\begin{array}{c} t(n)\,u^z(n) \\ u^z(n-1)
\end{array} \right)
\;=\;
{\cal T}^z(n,k)\;
\left(\begin{array}{c}t(k)\, u^z(k) \\ u^z(k-1)
\end{array} \right)
\mbox{ . }
\end{equation}

\noindent Note that the transfer matrices satisfy the transitivity
relation ${\cal T}^z(n,k)={\cal T}^z(n,m){\cal T}^z(m,k)$. A direct
inductive argument then shows that, for $\zeta\in\CC$,

\begin{equation}
\label{eq-transferid}
{\cal T}^{z+\zeta}(n,k)
\;=\;
{\cal T}^z(n,k)
-\zeta\,
\sum_{l=k}^{n-1}\,
{\cal T}^{z+\zeta}(n,l+1)\;
\frac{1}{t(l)}\,
\left(
\begin{array}{cc} 1 & 0 \\
0 & 0 
\end{array}
\right)\,
{\cal T}^{z}(l,k)
\mbox{ . }
\end{equation}

\noindent Taking the norm of (\ref{eq-transferid}), estimating the
r.h.s. and consequently taking the supremum over $0 \leq k \leq n \leq m$
leads to the following perturbative result which in a slightly different form is given in \cite{s} (see also \cite{DT}.)

\begin{lemma}
\label{lem-transferperturb}
Suppose
$$
\sup_{0\leq k\leq n\leq m}\;\|{\cal T}^{z}(n,k)\|\;\leq \;C
\qquad
D\;=\;\sup_{0\leq l\leq m-1}\frac{1}{|t_l|}
\mbox{ . }
$$

\noindent Then, as long as $CD|\zeta |m<1$, 

$$
\sup_{0\leq k\leq n\leq m}\;\|{\cal T}^{z+\zeta}(n,k)\|
\;\leq\; 
\frac{C}{1-CD|\zeta|m}
\mbox{ . }
$$

\end{lemma}

\subsection{Free Pr{\"u}fer variables}
\label{sec-pruefer}

Let now $E\in\RR$ and $u^E$ be given by (\ref{eq-eigenstates}).
The free Pr{\"u}fer phases $\theta^{0,E}(n)$ and amplitudes
$R^{0,E}(n)>0$ are now defined by

\begin{equation}
\label{eq-freeprufer}
R^{0,E}(n) \,
\left( \begin{array}{c} \cos (\theta^{0,E}(n))
\\ \sin(\theta^{0,E}(n)) \end{array} \right)
\;=\;
\left( \begin{array}{c} t(n)
u^E(n) \\ u^E(n-1) \end{array} \right)
\mbox{ , }
\end{equation}

\noindent the above initial conditions as well as

$$
-\frac{\pi}{2}  < \theta^{0,E}(n+1) - \theta^{0,E}(n)
< \frac{3\pi}{2}
\mbox{ . }
$$

\noindent Note that the $\theta^0$-dependence of the Pr{\"u}fer
variables is suppressed.

\begin{lemma}
\label{lem-Prueferderiv}
\begin{equation}
\label{eq-Prueferderiv}
R^{0,E}(n)^2\, {\partial_E} \,\theta^{0,E}(n)
\;=\;
\left\{
\begin{array}{cc}
\sum_{l=0}^{n-1} u^E(l)^2 & \mbox{if }\;\;n>0\mbox{ , }
\\
&
\\ - \sum_{l=n}^{-1} u^E(l)^2 & \mbox{if }\;\; n < 0 \mbox{ . }
\end{array}
\right.
\end{equation}
\end{lemma}

\noindent {\bf Proof:} From the recurrence relation
(\ref{eq-eigenstates}) and the definition of
$\theta^{0,E}(n)$ one gets

$$
\cot(\theta^{0,E}(n))
\; =\;
-t^2(n-1) \tan(\theta^{0,E}(n-1)) + v(n-1) - E
\mbox{ . }
$$

\noindent Differentiation leads to

$$
{\partial_E} \theta^{0,E}(n)
\;= \;
\frac{t^2(n-1) \sin^2(\theta^{0,E}(n))}{
\cos^2(\theta^{0,E}(n-1))} \;
{\partial_E}
\theta^{0,E}(n-1) + \sin^2(\theta^{0,E}(n))
\mbox{ . }
$$

\noindent Multiplying with $R^{0,E}(n)^2$ and using the
definition of $R^{0,E}(n)$ and $\theta^{0,E}(n)$ gives

\begin{equation}
\label{eq-intermed}
R^{0,E}(n)^2
{\partial_E} \theta^{0,E}(n)
\;= \;
R^{0,E}(n-1)^2
{\partial_E} \theta^{0,E}(n-1) + u^E(n-1)^2
\;.
\end{equation}

\noindent The above deduction of (\ref{eq-intermed}) has used that
$u^E(n-1) \not= 0$. If $u^E(n-1)=0$, then one may deduce
(\ref{eq-intermed}) in a similar way from
$$
\tan(\theta^{0,E}(n))
\;= \;
\frac{\cot(\theta^{0,E}(n-1))}{-t^2(n-1) + (v(n-1)-E)
\cot(\theta^{0,E}(n-1))}
\mbox{ . }
$$

\noindent The lemma now follows by iterating (\ref{eq-intermed}).
\hfill $\Box$

\vspace{.2cm}

Note in particular that (\ref{eq-Prueferderiv}) implies that
$\partial_E \theta^{0,E}(n)$ is strictly positive for $n\ge
2$ and strictly negative for $n\le -2$. Furthermore, it follows
from elementary considerations for transfer matrices that there
are constants $C_1$ and $C_2$ such that

\begin{equation}
\label{eq-derivbound}
0 \;<\; C_1\; \le \;
\left| {\partial_E}\, \theta^{0,E}(n) \right|
\;\le\; C_2\; < \;\infty
\mbox{ , }
\end{equation}

\noindent where $C_1$ and $C_2$ can be uniformly bounded away from
$0$ and $\infty$ as long as $|n|\geq 2$ and the quantities $|n|$,
$E$ and $\max_{|k|\le |n|} \{ |v(k)|, t(k), 1/t(k)\}$ remain
bounded (where only the lower bound requires $|n|\ge 2$).

\vspace{.2cm}

Let $\Pi_N$ be the projection on $\ell^2(\{0,\ldots,N-1\})$ and
denote the associated finite-size Jacobi matrix by $H_N=\Pi_N H\Pi_N$.
As $H_N$ has Dirichlet boundary conditions, let us choose
$u^E(-1)=0$ and $t(0)u^E(0)=1$ as initial conditions in the recurrence
relation (\ref{eq-eigenstates}). This corresponds to an initial
Pr{\"u}fer phase $\theta^0=0$. The formal solution $u^E$ then gives an
eigenvector (and $E$ is an eigenvalue of $H_N$) if and only if
$t(N)u^E(N)=R^{0,E}(N)\cos(\theta^{0,E}(N))=0$, that is
$\theta^{0,E}(N)=\frac{\pi}{2}\mod \pi$ (note herefore that $u^E(0)\neq 0$
for any eigenvector of $H_N$).

\vspace{.2cm}

One checks iteratively for all $n\geq 0$ that $u^E(n) > 0$ for $E$
sufficiently close to $-\infty$ and $\lim_{E\to -\infty} u^E(n-1)/u^E(n) =
0$. This and the definition of the Pr\"ufer phases implies that $\lim_{E\to
-\infty} \theta^{0,E}(n)=0$ for all $n\geq 0$, which one uses for $n=N$.
As $\theta^{0,E}(N)$ is
monotone increasing in $E$, it follows that the $j$th eigenvalue
$E_j$ of $H_N$ (counted from below $E_1<E_2<\ldots
<E_N$) satisfies
\begin{equation}
\label{eq-eigencount}
\theta^{0,E_j}(N)
\;= \;
\frac{\pi}{2} + \pi(j-1)
\mbox{ , }
\qquad
\theta^0\;=\;0
\mbox{ . }
\end{equation}

\noindent This oscillation theorem implies immediately:

\begin{equation}
\label{eq-oscithm}
\left| \frac{1}{\pi}\;\theta^{0,E}(N) \;-\;
\#\;\left\{\mbox{negative eigenvalues of}\;\;(H_N-E)\,
\right\}
\right|
\; \leq\;
\frac{1}{2}
\mbox{ . }
\end{equation}

\subsection{Modified Pr{\"u}fer variables}
\label{sec-modprufer}

Let us fix $M\in SL(2,\RR)$.  Set  $e_\theta=\left( \begin{array}{c}
\cos (\theta) 
\\ \sin(\theta) \end{array} \right)$. Define a
smooth function $m:\RR\to\RR$ with 
$m(\theta+\pi)=m(\theta)+\pi$ and
$0 < C_1 \le  m' \le C_2 < \infty$, by

$$
r(\theta)e_{m(\theta)}=Me_{\theta},
\qquad
r(\theta)>0
\mbox{ , }
\qquad
m(0)\in[-\pi,\pi)
\mbox{ . }
$$

\noindent Then the $M$-modified Pr{\"u}fer variables
$(R^{E}(n),\theta^{E}(n))\in \RR_+\times\RR$ for initial condition
$\theta^E(0)=\theta=m(\theta^{0})$
are given by
\begin{equation}
\label{eq-prufer1}
\theta^E(n)\;=\; m(\theta^{0,E}(n)) \mbox{ , }
\end{equation}
\noindent and
\begin{equation}
\label{eq-prufer2}
\left(\begin{array}{c} R^{E}(n) \cos
(\theta^{E}(n))
\\
R^{E}(n) \sin(\theta^{E}(n))
\end{array}
\right)
\;=\;
M \left( \begin{array}{c} t(n)\, u^E(n) \\ u^E(n-1)
\end{array} \right) 
\mbox{ , }
\end{equation}

\noindent where the dependence on the the initial phase is again
suppressed. 
Bounds of the form (\ref{eq-derivbound}) also hold for an $M$-modified
Pr{\"u}fer phase $\theta^E(n)$ because $\theta^E(n) = m(\theta^{0,E}(n))$
leads to
$(\min m') |\partial_E \theta^{0,E}| \le |\partial_E \theta^E|
\le (\max m') |\partial_E \theta^{0,E}|$.
Furthermore, as $|\theta^{E}(n)-\theta^{0,E}(n)|\leq 2\pi$,
(\ref{eq-oscithm}) implies that for the choice $\theta = m(0)$

\begin{equation}
\label{eq-oscithm2}
\left| \frac{1}{\pi}\;\theta^{E}(N) \;-\;
\#\;\{\mbox{negative eigenvalues of }\;\;(H_N-E)\,\} \right| \;\leq\;
\frac{5}{2}
\mbox{ . }
\end{equation}

\vspace{.2cm}

The goal to have in mind when choosing $M$ is to make the
$M$-modified transfer matrices as simple as possible so that the
$M$-modified Pr{\"u}fer variables are easy to calculate. Whenever $E$
is in the spectrum, the most simple
matrix to obtain is a rotation. Anything close to it can then be
treated by perturbation theory. This is the strategy followed for the
random polymer model below where $M$ is chosen as in (\ref{eq-tric})

\vspace{.2cm}

\noindent {\bf Example:} Let us consider an
$L$-periodic Jacobi matrix $H$. If $E\in\RR$ is in the interior of the 
spectrum of
$H$, there exists a matrix $M$ (depending on $E$, of course) such that
$MT^E(L)M^{-1}=R_\eta$ where $R_\eta$ is the rotation by an angle
$\eta=\eta(E)$ obtained in accordance with the definition (\ref{eq-prufer1}). 
The $M$-modified Pr{\"u}fer variables are then simply given by
$(R^{E}(kL),\theta^{E}(kL))=(1,k\eta)$ and the IDS is 
$\Nn(E)=\eta (E)/(L\pi)$.

\subsection{Covariant Jacobi matrices}
\label{sec-covJac}

Let $(\Omega,T,\ZZ,\PP)$ be a compact space $\Omega$, endowed with a
$\ZZ$-action $T$ and a $T$-invariant and ergodic probability measure
$\PP$. For a function $f\in
L^1(\Omega,\PP)$, let us denote $\EE(f(\omega))=\int
d\PP(\omega)\,f(\omega)$.
A strongly continuous family
$(H_\omega)_{\omega\in\Omega}$ of two-sided tridiagonal, self-adjoint
matrices on
$\ell^2(\ZZ)$ is
called covariant if the covariance relation
$UH_\omega U^*=H_{T\omega}$ holds where $U$ is the translation on
$\ell^2(\ZZ)$.
$H_\omega$ is characterized by two sequences $(t_\omega(n))_{n\in\ZZ}$
and $(v_\omega(n))_{n\in\ZZ}$ such that
(\ref{eq-polymerHam}) holds.

\vspace{.2cm}

The IDS at energy $E\in\RR$ of the family
$(H_\omega)_{\omega\in\Omega}$
can $\PP$-almost surely be defined by \cite{PF}

\begin{equation}
\label{eq-DOS}
\Nn(E)
\;=\;
\lim_{N\to\infty}\;\frac{1}{N}
\;\mbox{Tr}(\chi_{(-\infty,E]}(\Pi_NH_\omega\Pi_N))
\mbox{ , }
\end{equation}

\noindent while the Lyapunov exponent $\gamma(E)$ for $E\in\RR$
is $\PP$-almost surely given by the formula

$$
\gamma(E)
\;=\;
\lim_{N\to\infty}\frac{1}{N}
\log\left(\left\|{\cal T}^E_\omega(N,0)
\right\|\right)
\mbox{ , }
$$

\noindent where the transfer matrix ${\cal T}^E_{\omega}(N,0)$ from
site $0$ to $N$ is defined as in Section~\ref{sec-transfer}. Both the 
IDS and the Lyapunov
exponent are self-averaging quantities, notably
an average over $\PP$ may be introduced before taking the
limit without changing the result \cite{PF}.

\vspace{.2cm}

For each $H_\omega$ let
$(R^{E}_\omega(n),\theta^{E}_\omega(n))$ denote the associated
$M$-modified Pr{\"u}fer variables with some initial condition, 
then according to (\ref{eq-oscithm2})

\begin{equation}
\label{eq-IDS}
\Nn(E)
\;=\;
\lim_{N\to\infty}\;\frac{1}{\pi}\,\frac{1}{N}
\;\EE\left(\theta^E_\omega(N)\right)
\mbox{ . }
\end{equation}

While it is readily seen that $\gamma(E) \ge \lim^+_{N\to \infty}
\frac{1}{N} \EE(\log(R^E_{\omega}(N))$, one may in general not get
equality here as demonstrated by a counterexample in
Section~\ref{sec-polymers}. This is due to the dependence of
$R_{\omega}^E(N)$ on the initial phase $\theta$. The next lemma solves
this problem by (continuously) averaging over $\theta$.

\begin{lemma}
\label{lem-gamma}
For $E\in\RR$ and
any continuous {\rm (}{\sl i.e.} non-atomic{\rm )} measure $\nu$ on
$\RR P(1)= [0,\pi)$

\begin{equation}
\label{eq-LE2}
\gamma(E)
\;=\;
\lim_{N\to\infty}\;\frac{1}{N}
\;\int d\nu(\theta)
\;\EE\left(\log(R^E_\omega(N))\right)
\mbox{ . }
\end{equation}

\end{lemma}

\noindent {\bf Proof:} 
As
$\|{\cal T}^E_{\omega}(N,0) e_{m^{-1}(\theta)}\| = \|Me_{m^{-1}(\theta)}\|
\|M^{-1} e_{\theta_{\omega}^E(N)}\| R_{\omega}^E(N)$, a change of
variables and elementary estimates show that it is sufficient to
show that $\gamma(E)$ is equal to 

$$
\lim_{N\to\infty} 
\frac{1}{N} 
\int d\nu(\theta) \;
\EE\,(\log(\|{\cal T}^E_{\omega}(N,0) e_\theta\|))
$$ 

\noindent for any
continuous probability measure $\nu$. This is easy to see if
$\gamma(E)=0$, thus we now assume that $\gamma(E)>0$. 
Suppose the
contrary, that is there exists a $\nu$ such that

$$
\lim_{N\to\infty}
\frac{1}{N} 
\int d\nu(\theta)\;
\EE\,(\log(\|{\cal T}^E_{\omega}(N,0) e_{\theta}\|)) 
\;<\; \gamma(E)
\mbox{ . }
$$

\noindent By Fatou's lemma this implies that $\int
d\nu(\theta)\EE(\lim_{N\to\infty}\frac{1}{N} \log(\|{\cal T}^E_{\omega}(N,0)
e_{\theta}\|)) < \gamma(E)$. Because for a.e. $\omega$ the limit inside of the
expectation is equal to either $\gamma(E)$ or $-\gamma(E)$ by
Oseledec's Theorem, there has to exist a set
$\Ee\subset[0,\pi)\times\Omega$ of positive $\nu\otimes\PP$-measure
such that $\lim_{N\to\infty}\frac{1}{N}\log(\|{\cal T}^E_{\omega}(N,0)
e_{\theta}\|)$ is equal to $-\gamma(E)$ for all
$(\theta,\omega)\in\Ee$. 
Hence there exists an $\omega$ such that the set $\{\theta\in[0,\pi)\,|\, E
\mbox{ eigenvalue of }H_\omega(\theta)\}$ has positive $\nu$-measure
where $H_\omega(\theta)$ is the half-line operator with
$\theta$-boundary condition. 
As $\nu$ is continuous, this set has to contain 
at least two distinct points. This is in contradiction
to the fact that the difference equation $H_\omega u=Eu$ has, up to
constant multiples, at most
one square-summable solution at $+\infty$.
\hfill $\Box$

\section{Asymptotics of IDS and Lyapunov exponent}
\label{sec-Pruefer}

Generalizing the strategy suggested by Pastur and Figotin \cite{PF},
this chapter is devoted to the calculation of
the asymptotics for the IDS and the Lyapunov exponent near the
critical energy of a random polymer model, that is the proof of
Theorems \ref{theo-Lyapasym} and \ref{theo-DOSasym}.
The techniques of \cite{CS} allow to treat also the case
of strongly mixing (instead of random) configurations of polymer
chains giving similar formulas, containing a
correction factor given by the Fourier transform of the correlation
function. No further details are given here concerning this generalization.

\subsection{Random polymer chains}
\label{sec-polymers}

For sake of completeness, let us briefly indicate how to construct
$(\Omega,T,\ZZ,\PP)$ for the random polymer Hamiltonians
defined in Chapter
\ref{sec-res}. Let $\Omega_0$ be the Tychonov space of two-sided
sequences of signs. Set
$\Omega_\pm=\{\omega\in\Omega_0\,|\,\omega_0=\pm\} \times
\{0,\ldots,L_\pm-1\}$ and $\Omega=\Omega_+\cup\Omega_-$.
Now $T:\Omega\to\Omega$ is defined by

$$
T(\omega,l)
\;=\;
\left\{
\begin{array}{ccc}
(\omega,l+1) & & \mbox{if } \;l< L_{\omega_0}-1 \mbox{ , } \\
& & \\
(T_0\omega,0) & & \mbox{if } \;l= L_{\omega_0}-1 \mbox{ , }
\end{array}
\right.
$$

\noindent where $T_0$ is the left shift on $\Omega_0$. Now for any
set $A_{\pm}\subset\Omega_0$ of codes all having $\omega_0=\pm$,
one sets for all $l\in\{0\ldots L_\pm-1\}$

$$
\PP(\{(\omega,l)\in \Omega\,|\,\omega\in A_\pm\})
\;=\;\frac{\PP_0(A_\pm)}{\langle L_\pm\rangle}\,
\mbox{ , }
$$

\noindent where $\PP_0$ is the Bernoulli measure on $\Omega_0$. It
can then be verified that $\PP$ is invariant and ergodic (the
latter by mimicking the proof for $(\Omega_0,T_0,\PP_0)$). Random
hopping terms and potential are then given by $t_{(\omega,l)} = (\ldots,
\hat{t}_{\omega_0}, \hat{t}_{\omega_1}, \ldots)$ and $v_{(\omega,l)} =
(\ldots, \hat{v}_{\omega_0}, \hat{v}_{\omega_1}, \ldots)$ with choice
of origin $t_{(\omega,l)}(0)=\hat{t}_{\omega_0}(l)$
and $v_{(\omega,l)}(0)=\hat{v}_{\omega_0}(l)$. This leads to the
covariant family $(H_{(\omega,l)})_{(\omega,l)\in\Omega}$ of Jacobi
matrices. It is this family which is refered to as $(H_{\omega})$ in 
Section~\ref{sec-res} and, in particular, in
Theorems~\ref{theo-Lyapasym} to \ref{theo-DOSdeviations}.

\vspace{.2cm}

According to Section~\ref{sec-covJac} the Lyapunov exponent satisfies

\begin{equation}
\label{eq-ranlyapunov1}
\gamma(E)
\; = \;
\lim_{N\to\infty}
\frac{1}{N} 
\;\EE\,\left( \log
\left( 
\|{\cal T}_{(\omega,l)}^E(N,0)\|\right) \right) 
\; = \;
\lim_{N\to\infty}
\frac{1}{N}
\log (\|{\cal T}^E_{(\omega,l)}(N,0)\|)
\mbox{ , }
\end{equation}

\noindent for $\PP$-a.e.\ $(\omega,l)\in\Omega$. Here $\EE = \int
d\PP$. On the other hand, there is also a Lyapunov exponent associated with
random products of the unimodular matrices $T^E_{\pm}$:

\begin{equation}
\label{eq-ranlyapunov2}
\gamma_0(E)
\; = \;
\lim_{k\to\infty} 
\frac{1}{k}\;
\EE_0 \left( \log \left( \|T_{\omega}^E(k,0)\|\right) \right) 
\; = \;
\lim_{k\to\infty} \frac{1}{k}\, \log 
\left(\|T_{\omega}^E(k,0)\|\right)
\mbox{ , }
\end{equation}

\noindent for $\PP_0$-a.e.\ $\omega\in \Omega_0$ and $\EE_0 = \int d\PP_0$.
To compare $\gamma(E)$ and $\gamma_0(E)$, let $\tilde{\Omega}_0$ be
the full measure set of those $\omega \in \Omega_0$ such that
(\ref{eq-ranlyapunov2}) holds and also $\sum_{l=0}^{k-1}
L_{\omega_l}/k \to \langle L_{\pm} \rangle$ as $k\to\infty$. For
$\omega\in \tilde{\Omega}_0$ it is easily seen that $\lim_{N\to\infty}
\frac{1}{N} \log (\|{\cal T}^E_{(\omega,0)}(N,0)\|) =
\gamma_0(E)/\langle L_{\pm} \rangle$. Since $\PP\{(\omega,0)\;|\;
\omega \in \tilde{\Omega}_0\} = 1/\langle L_{\pm} \rangle >0$, one
concludes from (\ref{eq-ranlyapunov1}) that

\begin{equation}
\label{eq-LEcompare}
\gamma(E) \;= \;
\frac{1}{\langle L_{\pm} \rangle}\; \gamma_0(E) 
\mbox{ . }
\end{equation}

\noindent While $\gamma_0$ is not defined through a covariant operator, it
follows by the same argument as in Lemma~\ref{lem-gamma} that for any
continuous measure $\nu$ on $[0,\pi)$

\begin{equation}
\label{eq-LE}
\gamma_0(E)
\; = \;
\lim_{k\to\infty} \frac{1}{k}
\int d\nu(\theta) \EE_0
\log \left( \|MT^E_{\omega}(k,0)M^{-1} e_{\theta}\|\right)
\mbox{ . }
\end{equation}

\noindent {\bf Counterexample:} The continuity condition on $\nu$ 
in Lemma~\ref{lem-gamma} cannot be
weakened as shows the following example. Consider the polymer model
with $L_\pm=3$, $t(l)=1$ for all $l\in \ZZ$ and
$\hat{v}_+=(\frac{1}{2},2,0)$ and 
$\hat{v}_-=(-\frac{1}{2},-2,0),$ and choose $M={\bf 1}$. For $E=0$ it
is easily seen that $T_{\pm}^0 e_{\pi/2} = \mp \frac{1}{2} e_{\pi/2}$
and thus $\frac{1}{k} \log \|T_{\omega}^0(k,0) e_{\pi/2}\| =
-\frac{1}{2}$ for all $\omega$ and $k$, while
$\gamma_0(0)=\frac{1}{2}$. Hence a
measure having an atom at $\theta=\frac{\pi}{2}$ will not satisfy
(\ref{eq-LE}). This also provides a counterexample to
Lemma~\ref{lem-gamma} with $(\Omega,T,\ZZ,\PP)$ as above. For this one
uses that the event
$\{\omega\,|\,v_{\omega}(0) = \hat{v}_{\omega_0}(0)\}$ has probability
$1/3$ in $\Omega$.

\subsection{Polymer phase shifts}
\label{sec-phaseshift}

For $M$ given by (\ref{eq-tric}), let the polymer action multipliers
$\rho_{\pm}^{\epsilon}(\theta)$ and the polymer phase shifts
$\Ss_{\epsilon,\pm}(\theta)$ be the $M$-modified Pr\"ufer amplitude and
phase for the $L_{\pm}$-periodic polymers with initial phase $\theta$
at $0$ and evaluated at $L_{\pm}$ (i.e.\ over a single polymer
$(\hat{t}_+, \hat{v}_+)$ and $(\hat{t}_-, \hat{v}_-)$,
respectively). By definition of the modified Pr\"ufer variables, 
this means

\begin{equation}
\label{eq-polymerphaseshift}
\rho_\pm^\epsilon(\theta)
e_{\Ss_{\epsilon,\pm}(\theta)}
\;=\;
M T^{E_c+\epsilon}_\pm M^{-1}
e_{\theta}
\end{equation}

\noindent for all $\theta \in \RR$. The iterated polymer phase shifts 
are then denoted by

$$
\Ss_{\epsilon,\omega}^{l+1}(\theta)
\;=\;
\Ss_{\epsilon,\omega_l}
(\Ss_{\epsilon,\omega}^l(\theta))
\mbox{ , }
\qquad
\Ss_{\epsilon,\omega}^0(\theta) \;= \;\theta
\mbox{ . }
$$

\noindent From (\ref{eq-tric}) it follows
that (independent of $\theta$) $\rho_{\pm}^0(\theta) = 1$ and
$\eta_{\pm} = \Ss_{0,\pm}(\theta)-\theta$, at least up to a multiple of
$2\pi$ which is hereby fixed. The former readily implies that $\gamma(E_c)=0$.
To study the Lyapunov exponent in a vicinity of $E_c$, iterate 
(\ref{eq-polymerphaseshift}) in order to deduce

\begin{equation}
\label{eq-exacttransfer}
\log \left( \left\| 
MT^{E_c+\epsilon}_{\omega}(N,0)M^{-1}e_{\theta}
\right\| \right)
\;=\; 
\sum_{l=0}^{N-1} \log \left( 
\rho_{\omega_l}^\epsilon
  (\Ss_{\epsilon,\omega}^l(\theta)) \right)
\mbox{ , }
\end{equation}

\noindent which combined with (\ref{eq-LEcompare}) and (\ref{eq-LE}) gives  

\begin{equation}
\label{eq-Lyappol}
\gamma(E_c+\epsilon)
\;=\;
\;\frac{1}{\langle L_\pm\rangle}\,
\lim_{N\to\infty}\;\frac{1}{N}\,\sum_{l=0}^{N-1}
\;
\int d\nu(\theta)\;
\EE_0\left(\log(\rho_{\omega_l}^\epsilon(\dyn^l(\theta)))\right)
\mbox{ . }
\end{equation}

\vspace{.2cm}

To also express the IDS in terms of the polymer phase shifts, let
$(n_{(\omega,l),k})_{k\in\ZZ}$ be the sequence of lower polymer nodes
for a given $(\omega,l)\in \Omega$, i.e.\ the integers determined by
$v_{(\omega,l)}(n_{(\omega,l),k}) = \hat{v}_{\omega_k}(0),$ for any choice of $\hat{v}.$ For $N\in
\NN$, let $n_{(\omega,l),k}$ be the polymer node closest to $N$. Since
$\Ss_{\epsilon,\omega}^k(\theta)-\theta$ is a rotation number for a
matrix which arises from $H_{(\omega,l),N}$ by a perturbation of rank
bounded by $C\max \{L_-,L_+\}$, it follows that
$|\theta_{(\omega,l)}^{E_c+\epsilon}(N) -
(\Ss_{\epsilon,\omega}^k(\theta) - \theta)| \le C \max \{L_-,L_+\}$
uniformly in $\theta$. Thus it follows from (\ref{eq-IDS}) that
$\Nn(E_c + \epsilon) = \lim_{N\to\infty} \frac{1}{\pi N}
(\Ss_{\epsilon,\omega}^k(\theta)-\theta)$ almost surely and in
expectation. Since $k/N \to 1/\langle L_{\pm} \rangle$ almost surely
as $N\to\infty$, this implies that

\begin{eqnarray}
\label{eq-IDSpol}
\Nn(E_c+\epsilon) 
& = & \frac{1}{\pi \langle L_{\pm} \rangle}\;
\lim_{k\to\infty}\;
\frac{1}{k} \;\EE_0(\Ss_{\epsilon,\omega}^k(\theta)-\theta) 
\nonumber \\
& = & \frac{1}{\pi \langle L_{\pm} \rangle}
\;\lim_{k\to\infty}\; 
\frac{1}{k}\; \sum_{l=0}^{k-1} 
\EE_0 \left(\Ss_{\epsilon,\omega_l}(\Ss_{\epsilon,\omega}^{l}(\theta)) 
- \Ss_{\epsilon,\omega}^{l}(\theta)\right)
\mbox{ . }
\end{eqnarray}

\subsection{Calculation of phase shifts and
action multipliers}
\label{sec-multpha}

The aim of this paragraph is to calculate the polymer phase shifts and action
multipliers needed in (\ref{eq-IDSpol}) and
(\ref{eq-Lyappol}) in terms of the transmission and reflection coefficients
defined in (\ref{eq-reflec}).
Because $\mbox{det}(MT^{E_c+\epsilon}_{\pm}M^{-1})=1$, these
coefficients satisfy

$$
|a_\pm^\epsilon|^2-|b_\pm^\epsilon|^2\;=\;1
\mbox{ . }
$$

\noindent A further short calculation shows that

\begin{equation}
\label{eq-complexrho}
\rho_\pm^\epsilon(\theta)^2
\;=\;
1+
2\,\Re e\left(
a_\pm^\epsilon {b}_\pm^\epsilon \,
e^{2\imath\theta}\right)
+
2|b_\pm^\epsilon|^2
\mbox{ , }
\end{equation}

\noindent and

$$
e^{\imath(\Ss_{\epsilon,\pm}(\theta)\,-\,\theta)}
\;=\;
\frac{a_\pm^\epsilon  + \overline{b_\pm^\epsilon}
e^{-2\imath\theta}}{
\left|a_\pm^\epsilon  + \overline{{b}_\pm^\epsilon} e^{-2\imath\theta}\right|}
\mbox{ . }
$$

\vspace{.2cm}

Now using the phase $\eta^\epsilon_\pm$  of
$a_\pm^\epsilon$,

\begin{equation}
\label{eq-expan1}
a_\pm^\epsilon
\;=\;
e^{\imath \eta^\epsilon_\pm}+\Oo(|b_\pm^\epsilon|^2)
\mbox{ . }
\end{equation}

\noindent This leads to the following expansions:

\begin{equation}
\label{eq-expan3}
\log(\rho_\pm^\epsilon(\theta)^2)
\;=\;
2\,\Re e\left(
a_\pm^\epsilon {b}_\pm^\epsilon
e^{2\imath\theta}\right)
+|b_\pm^\epsilon|^2
-
\Re e\left(
(a_\pm^\epsilon {b}_\pm^\epsilon)^2
e^{4\imath\theta}\right)
+\Oo\left(|b_\pm^\epsilon|^3\right)
\mbox{ , }
\end{equation}

\noindent and

\begin{equation}
\label{eq-expan4}
e^{2\imath(\Ss_{\epsilon,\pm}(\theta)-\theta)}
\;=\;
e^{2\imath\eta^\epsilon_\pm}
+
\overline{{b}_\pm^\epsilon}e^{\imath\eta^\epsilon_\pm}
\,e^{-2\imath\theta}
-
{b}_\pm^\epsilon e^{3\imath\eta^\epsilon_\pm}
\,e^{2\imath\theta}
+\Oo\left(|b_\pm^\epsilon|^2\right)
\mbox{ . }
\end{equation}

\subsection{Oscillatory sums}

\begin{proposi}
\label{prop-oscisums}
Let $c_\pm\in\CC$, $j=1,2$, and set

$$
I^j_N(\theta,\epsilon)\;=\;
\EE_0(I^j_{\omega,N}(\theta,\epsilon))
\mbox{ , }
\qquad
I^j_{\omega,N}(\theta,\epsilon)
\;=\;
\sum_{l=0}^{N-1}
\;c_{\omega_{l}}\,e^{2\imath j \Ss_{\epsilon,\omega}^l(\theta)}
\mbox{ . }
$$

\noindent 
Let $\epsilon$ be sufficiently small.
If $\langle e^{2\imath j \eta_\pm}\rangle\neq 1$, then
$I^j_N(\theta,\epsilon)=
\Oo(N|b^\epsilon_\pm|,\,1)$. If
$\langle e^{2\imath j\eta_\pm}\rangle\neq 1$ for both $j=1,2$,

$$
I^1_N(\theta,\epsilon)
\;=\;
N\;\langle c_\pm\rangle\;
\frac{
\langle \overline{{b}^\epsilon_\pm}e^{\imath\eta^\epsilon_\pm}\rangle
}{
1-\langle e^{2\imath\eta^\epsilon_\pm}\rangle}
+\Oo(N |b^\epsilon_\pm|^2,\,1)
\mbox{ . }
$$

\end{proposi}

\noindent {\bf Proof:}
Since
$\Ss_{\epsilon,\omega}^{l+1}(\theta)=\Ss_{\epsilon,\omega_l}
(\Ss_{\epsilon,\omega}^{l}(\theta))$
and $\Ss_{\epsilon,\omega}^{l}(\theta)$
is independent of
$\omega_l$, one gets

\begin{equation}
I^1_N(\theta,\epsilon)=
\langle e^{2\imath \eta^\epsilon_\pm}\rangle\,
I^1_{N-1}(\theta,\epsilon)+
\langle c_\pm\rangle\, e^{2\imath \theta}
+
\langle c_\pm\rangle\sum_{l=1}^{N-1}
\EE_0\left(e^{2\imath \Ss_{\epsilon,\omega_l}
(\Ss_{\epsilon,\omega}^{l}(\theta))}
-e^{2\imath (\eta^\epsilon_{\omega_l}+
\Ss_{\epsilon,\omega}^{l}(\theta))}\right)
.
\label{eq-devel}
\end{equation}

\noindent Equation (\ref{eq-expan4}) shows that
$e^{2\imath \dynl(\theta)}
-e^{2\imath (\eta^\epsilon_{\omega_l}+\theta)}=\Oo(|b^\epsilon_\pm|)$.
As $I^1_N(\theta,\epsilon)=I^1_{N-1}(\theta,\epsilon)+\Oo(1)$
and $\langle e^{2\imath \eta^\epsilon_\pm}\rangle\neq 1$ by hypothesis, one can
solve for $I^1_N(\theta,\epsilon)$ which directly implies that
$I^1_N(\theta,\epsilon)= \Oo(N|b^\epsilon_\pm|\,1)$.
Along the same lines, $I^2_N(\theta,\epsilon)=
\Oo(N|b^\epsilon_\pm|,\,1)$.
Now insert the expansion
(\ref{eq-expan4}) in (\ref{eq-devel}).
Due to the above, the oscillatory
terms in that formula are then of order 
$\Oo(N|b_\pm^\epsilon|^2)$.
Thus only the
non-oscillatory term on the r.h.s. of (\ref{eq-expan4}) gives a
contribution to the leading order. 
\hfill $\Box$

\subsection{Asymptotics of the IDS}

\noindent {\bf Proof} of Theorem \ref{theo-DOSasym}:
Formula (\ref{eq-expan4}) leads to $\Ss_{\epsilon,\pm}(\theta) -
\theta = \eta_{\pm} + \epsilon d_{\pm} - \epsilon \,\Im m (c_{\pm}
{e}^{2\imath\theta}) + \Oo(\epsilon^2)$, where $d_{\pm} =
\left.(\partial_{\epsilon} \eta_{\pm}^{\epsilon})\right|_{\epsilon=0}$
and $c_{\pm} = \left.(\partial_{\epsilon}
  b_{\pm}^{\epsilon})\right|_{\epsilon =0}
{e}^{\imath\eta_{\pm}}$. Inserting this in (\ref{eq-IDSpol}) yields

$$
\Nn(E_c+\epsilon) = 
\frac{1}{\pi \langle L_{\pm} \rangle} 
\left(
  \langle \eta_{\pm} \rangle + \epsilon \,\langle d_{\pm} \rangle 
- \epsilon \lim_{k\to\infty} \frac{1}{k} \;\Im m \;\EE_0
\left( \sum_{l=0}^{k-1}
    c_{\omega_l} {e}^{2\imath S_{\epsilon,\omega}^{l}(\theta)}
\right) 
+ \Oo(\epsilon^2) 
\right) 
\mbox{ . }
$$

\noindent By Proposition~\ref{prop-oscisums}, the expectation of the
oscillatory sum is of order $\Oo(k|b_{\pm}^{\epsilon}|,1)$ and thus

\begin{equation}
\label{eq-IDSexp}
\Nn(E_c+\epsilon)
 = \frac{1}{\pi \langle L_{\pm} \rangle} \left( \langle
  \eta_{\pm} \rangle + \epsilon \langle d_{\pm} \rangle +
  \Oo(\epsilon^2) \right) \mbox{ . }
\end{equation}

\noindent Setting $p_+=1$ and $p_+=0$ yields that in particular
  $\Nn_{\pm}(E_c+\epsilon) = \frac{1}{\pi L_{\pm}} (\eta_{\pm} + \epsilon
  d_{\pm} + \Oo(\epsilon^2))$, allowing to identify $\Nn_{\pm}(E_c) =
  \eta_{\pm}/\pi L_{\pm}$ and $\Nn_{\pm}'(E_c) = d_{\pm}/\pi
  L_{\pm}$. Using this to insert for $\eta_{\pm}$ and $d_{\pm}$ in
  (\ref{eq-IDSexp}) completes the proof. \hfill $\Box$

\vspace{.2cm}

\subsection{Asymptotics of the Lyapunov exponent}
\label{subsec-Lyap}

\noindent {\bf Proof} of Theorem \ref{theo-Lyapasym}:
Replacing (\ref{eq-expan3}) into (\ref{eq-Lyappol}) shows that
$2\langle L_\pm\rangle\,\gamma(E_c+\epsilon)$ is, up to corrections of
order $\Oo(|b_\pm^\epsilon|^3)$,
equal to the $\nu$-average of
$$
\langle |b_\pm^\epsilon|^2\rangle
+
2\,\Re e\left(\langle a_\pm^\epsilon b_\pm^\epsilon\rangle
\lim_{N\to\infty}\frac{1}{N}\sum_{l=0}^{N-1}
\EE_0\left(e^{2\imath\dyn^l(\theta)}\right)\right)
-
\Re e\left(
\langle (a_\pm^\epsilon b_\pm^\epsilon)^2\rangle
\lim_{N\to\infty}\frac{1}{N}\sum_{l=0}^{N-1}
\EE_0\left(e^{4\imath\dyn^l(\theta)}\right)\right)
.
$$
By Proposition \ref{prop-oscisums}, the first oscillatory sum has a
contribution of the order $\Oo(|b_\pm^\epsilon|^2)$ (which is given
there) while the second oscillatory sum is of
order $\Oo(|b_\pm^\epsilon|^3)$ and can hence be neglected.
Therefore one obtains

\begin{equation}
\label{eq-lyapres2}
\gamma(E_c+\epsilon)
\; =\;
\frac{1}{\langle L_\pm\rangle}\,
\left[\frac{1}{2}\;
\langle |b^\epsilon_\pm|^2\rangle
+\Re e
\left(
\frac{\langle b^\epsilon_\pm e^{\imath \eta^\epsilon_\pm}\rangle\;
\langle \overline{b^\epsilon_\pm} e^{\imath \eta^\epsilon_\pm}\rangle}{
1-\langle e^{2\imath \eta^\epsilon_\pm}\rangle}
\right)
\right]
+\Oo\left(|b^\epsilon_\pm|^3\right)
\mbox{ . }
\end{equation}

\noindent It can be directly verified that the given leading order
term vanishes if either $p_+=0$ or $p_+=1$, which also follows
from the fact that in this case $H_{\omega}$ is a periodic Jacobi
matrix, whose Lyapunov exponent vanishes in the interior of its
spectral bands. Next rewrite (\ref{eq-lyapres2}) as a fraction
with common denominator $|1- \langle e^{2\imath
\eta^\epsilon_\pm}\rangle|^2$. Since $p_-=1-p_+$, the numerator is a
polynomial of degree at most $3$ in $p_+$ vanishing at $p_+=0$ and
$p_+=1$. Elementary but lengthy algebra shows that moreover its
third derivative vanishes identically. Calculating the first order
derivative allows to conclude. \hfill $\Box$

\subsection{Comments}
\label{subsec-comments}

A random phase approximation consists in supposing that the incoming
phases $\Ss_{\epsilon,\omega}^{l}(\theta)$
in each summand of (\ref{eq-Lyappol}) and (\ref{eq-IDSpol}) 
is completely random, that is distributed according to the Lebesgue
measure. It can easily be checked that one actually obtains the correct
answers for the derivatives of both the IDS and the Lyapunov exponent at
the critical energy within this approximation. However, the lowest order
non-vanishing term in the Lyapunov exponent is
$\Oo(|b^\epsilon_\pm|^2)$ and the random phase approximation gives 
together with the expansion (\ref{eq-expan3}) that
$\gamma(E_c+\epsilon) \approx
\langle |b^\epsilon_\pm|^2\rangle / (2\,\langle L_\pm\rangle)$, namely
only the first term in (\ref{eq-lyapres2}). As we shall argue now, 
the second contribution is
due to the presence of correlations (or memory) in the family of
discrete time random dynamical systems
$(\dynpm,\RR P(1),\Omega,\PP)_{\epsilon\in\RR}$. 

\vspace{.2cm}

It is a result of Furstenberg \cite{Fur} (his hypothesis can be
checked here) that for each $\epsilon\neq 0$ (small enough)
there exists a unique invariant measure $\nu_\epsilon$
on $\RR P(1)$ satisfying

$$
\int d\nu_\epsilon(\theta) \,f(\theta)
\;=\;
\int d\nu_\epsilon(\theta)
\,\langle f(\dynpm(\theta))\rangle
\mbox{ , }
\qquad
f\in C(\RR P(1))
\mbox{ . }
$$

\noindent For $\epsilon=0$, one invariant measure is given by the
Lebesgue measure (it is unique if $\eta_+-\eta_-$ is irrational). 
For finite $\epsilon$, iteration of the invariance
property and the Proposition \ref{prop-oscisums} implies

$$
\int d\nu_\epsilon(\theta)\,e^{2\imath \theta}
\;=\;
\int d\nu_\epsilon(\theta)\;
\EE\left(\frac{1}{N}\sum_{l=0}^{N-1}
e^{2\imath \dyn^l(\theta)}\right)
\;=\;
\frac{
\langle \overline{b^\epsilon_\pm}e^{\imath\eta^\epsilon_\pm}\rangle
}{
1-\langle e^{2\imath\eta^\epsilon_\pm}\rangle}
+
\Oo(|b^\epsilon_\pm|^2)
\mbox{ . }
$$

\noindent Similarly $\int d\nu_\epsilon(\theta)\,e^{4\imath \theta}
=\Oo(|b^\epsilon_\pm|)$. These facts express the deviations of the
invariant measure from the Lebesgue measure and hence from the random phase
approximation. Moreover, for small enough $\epsilon\neq 0$ the
invariant measure $\nu_\epsilon$ is known to be H{\"o}lder continuous
\cite[p. 161]{BL} so that one can use it in (\ref{eq-Lyappol}). Hence

$$
\gamma(E_c+\epsilon)
\;=\;
\frac{1}{2}\;\frac{1}{\langle L_\pm\rangle}\,
\int d\nu_\epsilon(\theta)\;
\left\langle\log(\rho_\pm^\epsilon(\theta)^2))\right\rangle
\mbox{ . }
$$

\noindent Developing $\log(\rho_\pm^\epsilon(\theta)^2)$ as in
(\ref{eq-expan3}) then also leads an alternative proof of 
(\ref{eq-lyapres2}) and the
second contribution in (\ref{eq-lyapres2}) is indeed due to the
correlations as claimed above. Finally let us point out that higher
order terms in $\epsilon$ can readily be calculated, under adequate
(weak) hypothesis.

\vspace{.2cm}

\section{Large deviation estimates}
\label{sec-deviations}

Using elementary estimates on the boundary terms $M$ and $M^{-1}$ in
(\ref{eq-exacttransfer}), as well as (\ref{ab0}), 
and the expansions (\ref{eq-expan1}) and 
(\ref{eq-expan3}), one
obtains that for all $0\leq m\leq k\leq N$

\begin{equation}
\label{eq-lypdeloc}
\log
\left(
\left\|T^{E_c+\delta}_{\omega}(k,m)\right\|^2
\right)
\;=\;
2\delta\;\sup_{\theta\in [0,\pi)}
\,\Re e\,
\sum_{l=m}^{k-1}
{c}_{\omega_l} \,
e^{2\imath\Ss^{l}_{\delta,\omega}(\theta)}
+
\Oo(N\delta^2,1)
\mbox{ , }
\end{equation}

\noindent where $c_\pm={e}^{\imath\eta_\pm}(\partial_\delta 
b^\delta_\pm)|_{\delta=0}$. If
the order of the critical energy is $1$, then $c_\pm=\Oo(1)$.
In order to prove the delocalization
results, it is necessary to show that the l.h.s. of (\ref{eq-lypdeloc})
is of order 1 as long as $\Oo(N\delta^2)=1$. Therefore one needs to
show that sums like $I^1_{\omega,k}(\theta,\delta)$ 
defined in Proposition \ref{prop-oscisums} are with high probability
of order $\sqrt{N}$ for $\Oo(N\delta^2)=1$ and $|k|\leq N$.
These random Weyl sums can be thought of
as a discrete time (variable $N$)
correlated random walk in the complex plain,
the correlation being due to the presence of the
dynamics $\Ss_{\delta,\pm}$. For the present purposes, it is
sufficient to show that this sum actually
behaves as a random walk on adequate time scales.
Hence let us introduce, for every $\delta$, $\theta$,

\begin{equation}
\label{eq-badset}
{\Omega}^0_{N}(\alpha,\delta,\theta)
\;=\;
\left\{
\omega\in\Omega_0\,\left| \,\exists\;
k\leq N\mbox{ such that }
|I^1_{\omega,k}(\theta,\delta)|\geq N^{\alpha+1/2}
 \right\}\right.
\mbox{ . }
\end{equation}

\begin{theo}
\label{theo-corest}
If $|\langle e^{2\imath \eta_\pm}\rangle|< 1$ and $\alpha >0$, 
there exist constants $C_1$ and $C_2$ such that for all $\theta,$ $N$ and
$\delta$ with $N\delta^2 \le 1$:
\begin{equation}
\label{eq-badsetest}
{\PP}_0({\Omega}^0_{N}(\alpha,\delta,\theta))
\;\leq\;
C_1\,e^{-C_2N^{\alpha}}
\mbox{ . }
\end{equation}

\end{theo}

The proof of this estimate will be given in
Section~\ref{sec-corel}. First, let us deduce the following consequence:

\begin{theo}
\label{btt}
Let $|\langle {e}^{2i\eta_{\pm}} \rangle |<1$ and $\alpha>0$. Then
there are $c, c'>0$, $C<\infty$ such that for every $N\in \NN$, there
exists a set $\Omega_N(\alpha) \subset \Omega$ satisfying

$$
\PP(\Omega_{N}(\alpha))
\;=\;
\Oo(e^{-cN^{\alpha}})
\mbox{ , }
$$

\noindent and such that for every configuration $(\omega,l)$ in the
complementary set $\Omega_{N}(\alpha)^c=\Omega\backslash 
\Omega_{N}(\alpha)$, one has

$$
\left\|{\cal T}^{E_c+\delta+\imath\kappa}_{(\omega,l)}(k,m)\right\| 
\;\leq \;C
\mbox{ , }
$$

\noindent for all $0\leq m\leq k\leq N$ and all $|\delta| \leq N^{-\alpha
  -1/2}, \;|\kappa| \leq c'/N.$
 
\end{theo}

\noindent {\bf Proof:}
In order to estimate the norms of the transfer matrices using the Weyl
sums, note that for any $2\times 2$ matrix $A$,

\begin{equation}
\label{eq-norm}
\left\|A\right\|
\;=\; 
\sup_{\theta\in [0,\pi)}\left\|Ae_{\theta}\right\|
\;\leq\;
\sqrt{2}\max_{\theta =0, {\pi\over 2}}\left\|Ae_{\theta}\right\|
\mbox{ . }
\end{equation}

Set $\Omega_N^0 (\alpha,\delta)=\Omega_N^0 (\alpha,\delta,0)\cup
\Omega_N^0 (\alpha,\delta,{\pi\over 2})$. 
Then, combining  Theorem \ref{theo-corest} with (\ref{eq-lypdeloc})
and (\ref{eq-norm}) as well as the fact that  
$T(k,m)=T(k,0)T(m,0)^{-1}$, one
deduces that for all
$\omega\in{\Omega}_N^0(\alpha,\delta)^c$ with $|\delta | < N^{-\alpha -1/2}$, 
norms of the transfer-matrices ${T}^{E_c+\delta}_{\omega}(k,m)$ 
are uniformly bounded 
by a constant, not dependent on $\delta.$ 

Now let ${\Omega}_N(\alpha,\delta) = \{ (\omega,l)\in \Omega \left|
\;\omega \in {\Omega}^0_N(\alpha,\delta)\} \right.$. 
It follows that 

$$
\PP({\Omega}_N(\alpha,\delta)) 
\;\le\; \frac{L_++L_-}{\langle L_{\pm}
  \rangle}\; 
\PP_0({\Omega}^0_N(\alpha,\delta)) 
\;\le \;C_5
{e}^{-C_4N^{\alpha}}. 
$$

Elementary estimates (based on uniform bounds on norms of transfer
matrices over blocks of length no more than $\max L_{\pm}$) imply that
$$
\| {\cal T}^{E_c+\delta}_{(\omega,l)} (k,m)\| \;\le \;C'
\mbox{ , }
$$
\noindent for all $(\omega,l) \in \Omega_N(\alpha,\delta)^c$,
$|\delta| \le N^{-\alpha-1/2}$ and $0\leq m\leq k\leq N$ (in fact this
holds for $m,k$ up to the $N$-th polymer node). Set $\epsilon =
N^{-\alpha-1/2}$. The theorem then follows from
Lemma~\ref{lem-transferperturb}, by taking $\Omega_N(\alpha) =
\bigcup_{k=-N}^N \Omega_N(\alpha, k\epsilon/N)$. \hfill $\Box$

\subsection{Correlation bounds: Proof of Theorem \ref{theo-corest}}
\label{sec-corel}

\begin{lemma}
\label{lem-decay}
Let
$\kappa=|\langle e^{2\imath \eta_\pm}\rangle|< 1$. Then there exists a
centered complex random variable $X(\omega)$ depending on
$\omega_1,\ldots,\omega_r$ such that

$$
e^{2\imath {\cal S}_{\delta,\omega}^r(\theta)}
\;=\;
X(\omega)e^{2\imath\theta}
+\Oo(r\delta,\kappa^r)
\mbox{ . }
$$

\noindent Moreover, $|X(\omega)|$ is uniformly bounded by $2$.

\end{lemma}

\noindent {\bf Proof:}
Let us set 
$\kappa_r(\omega)=
\exp(2\imath \sum_{m=1}^{r}\eta_{\omega_m})$.
Note that 
$|\EE_0(\kappa_r(\omega))|=\kappa^r$.
Iteration of
$e^{2\imath \Ss_{\delta,\pm}(\theta)}=e^{2\imath (\eta_\pm+\theta)}
+\Oo(\delta)$ and centering the random variable
$\kappa_r(\omega)$ shows

$$
e^{2\imath {\cal S}_{\delta,\omega}^r(\theta)}
\;=\;
(\kappa_{r}(\omega)-\langle e^{2\imath\eta_\pm}\rangle^r)
e^{2\imath\theta}
+
\Oo(\kappa^r, \delta r)
\mbox{ , }
$$

\noindent as claimed.
\hfill $\Box$

\vspace{.2cm}

\noindent {\bf Proof} of Theorem \ref{theo-corest}:
Let $r$ be the smallest integer larger than
$\log(N^{-\alpha-1/2})/\log(\kappa)$.
Applying Lemma \ref{lem-decay} to
each term (except the first $r$ terms) of the sum
$I^1_{\omega,k}(\theta,\delta)$ shows that 

\begin{equation}
\label{eq-intermedest}
I^1_{\omega,k}(\theta,\delta)
\;=\;
\sum_{l=0}^{k-1}c_{\omega_{l+r}}\;X(T_0^l\,\omega)
\,e^{2\imath 
{\cal S}_{\delta,\omega}^l(\theta)}
+
\Oo(rk\delta,\,k{\kappa}^r,\,r)
\mbox{ . }
\end{equation}

\noindent 
(Here the identity ${\cal S}_{\delta,\omega}^{l+r}(\theta) =
{\cal S}_{\delta,T_0^l\omega}({\cal S}_{\delta,\omega}^l(\theta))$ was used. 
Recall moreover that $T_0^l\omega$ is the $l$-fold shift of $\omega$.)
Under the hypothesis of the theorem and because of the
choice of $r$, the error term in
(\ref{eq-intermedest}) is $\Oo(N^{1/2}\log N)$. Thus it is sufficient 
to prove probabilistic estimates of the
appearing sum, which will be denoted by $Z_k(\omega,\delta)$.

In order to decouple the correlations, divide $\{0,\dots,k-1\}$ in
$2R$ pieces $I_0, \ldots, I_{2R-1}$ of equal length $[k^{\alpha}]$,
where $R= [k/(2[k^{\alpha}])]$, i.e.\ $I_s = \{ s[k^{\alpha}], (s+1)
[k^{\alpha}]-1\}$, $s=0,\ldots,2R-1$. Here $[x]$ denotes the largest
integer smaller or equal to $x$. This excludes $ck^{\alpha}$ terms
which in the following can be absorbed in the error. Set for $j=0,1$:

$$
Z^j_R(\omega,\delta)
\;=\;
\sum^{R-1}_{s=0} Y_{2s+j}(\omega,\delta)
\mbox{ , }
\qquad
Y_s(\omega,\delta)
\;=\;\sum_{l\in I_s}
c_{\omega_{l+r+1}}\;X(T_0^l\omega)
\,e^{2\imath 
{\cal S}_{\delta,\omega}^l(\theta)}
\mbox{ . }
$$

\noindent Thus

\begin{equation}
\label{eq-corrdecoup}
Z_k(\omega,\delta)
\;=\;
Z^0_R(\omega,\delta) +
Z^1_R(\omega,\delta) + 
\Oo(k^{\alpha})
\mbox{ . }
\end{equation}
 
\noindent The random variable
$Y_s(\omega,\delta)$ satisfies uniformly
$|Y_s(\omega,\delta)|\leq c_1\, k^\alpha.$ 
If $\EE_s$ denotes the averaging procedure
(conditional expectation) over all random variables $\omega_l$ for
$l\geq s$, then Lemma~\ref{lem-decay} implies $\EE_{s[k^\alpha]+1}
(Y_s(\omega,\delta))=0$.

In the following estimates, real and imaginary parts of $Z^j_R(\omega,\delta)$
are treated separately, but in exactly the same way; hence one may
suppose that $Z^j_R(\omega,\delta)$ and all the summands therein are real. 
For $\lambda>0$ and
$\beta>0$, the Tchebychev and Cauchy-Schwarz inequalities imply

$$
\PP_0(\{\omega\in\Omega_0
\,|\,Z_k(\omega,\delta)>\lambda\})
\;\leq\;
e^{-\beta\lambda}\;
\EE_0(e^{\beta Z_k(\omega,\delta)})
\;\leq\;
e^{-\beta\lambda+C\beta k^{\alpha}}\;\max_{j=0,1}\;
\EE_0(e^{2\beta Z^j_R(\omega,\delta)})
\mbox{ . }
$$

Now if $-1\leq Y\leq 1$, by convexity
$2\,e^{\beta Y}\leq (1-Y)e^{-\beta}+(1+Y)e^\beta$. Thus if $Y$ is a
real centered random variable,

\begin{equation}
\label{eq-expexpbound}
\EE(e^{\beta Y(\omega)})
\;\leq\;
(e^{-\beta}+e^{\beta})/2
\;\leq\; 
e^{\beta^2/2}
\mbox{ . }
\end{equation}

\noindent One may assume that $[k^{\alpha}] > r-1$ and $k\geq
N^{\frac{1+\alpha}{2}}$ (otherwise it is trivially 
true that $|I_{\omega,k}^1(\theta,\delta)| < N^{\alpha+1/2}$). Thus
$Z^j_{R-1}(\omega,\delta)$ does not depend on the $\omega_l$ with $l
\ge (2(R-1)+j)[k^{\alpha}]-1$ and a rescaled
version of (\ref{eq-expexpbound}) can be iteratively applied to the 
conditional expectations, leading to

\begin{eqnarray*}
\EE_0(e^{2\beta Z^j_R(\omega,\delta)})
& \leq &
\EE_0\left(\EE_{(2(R-1)+j)[k^{\alpha}]-1}(e^{2\beta 
Y_{2(R-1)+j}(\omega,\delta)})
\,e^{2\beta Z^j_{R-1}(\omega,\delta)}\right) \\
& \leq &
e^{(2c_1k^\alpha\beta)^2/2}
\EE_0
\left(e^{2\beta Z^j_{R-1}(\omega,\delta)}\right) \\
& \leq &
e^{c_2 \beta^2 k^{1+\alpha}}
\mbox{ . }
\end{eqnarray*}

\noindent Choosing $\beta=\lambda/(2c_2 k^{1+\alpha})$ and
proceeding similarly for
$\{\omega\in\Omega_0\,|\,
Z_k(\omega,\delta)<-\lambda\}$ thus shows
(after recombining real and imaginary parts)

$$
\PP_0
(\{\omega\in\Omega_0\,
|\,|Z_k(\omega,\delta)|>\lambda\})
\;\leq\;
4\,e^{-{\lambda}^2/(4c_2\,k^{1+\alpha})+\frac{C\lambda}{2c_2k}}
\mbox{ . }
$$

\noindent Using this estimate for
$\lambda=N^{\alpha+1/2}$ and renormalizing the constants in order to
compensate for the error terms in (\ref{eq-intermedest}) 
as well as for summation over $k$ 
concludes the proof.
\hfill $\Box$

\subsection{Eigenvalue distribution in the metalic phase}
\label{sec-DOS}

\vspace{.2in}

\noindent {\bf Proof} of Theorem \ref{theo-DOSdeviations}: 
Let us fix  a configuration $(\omega,l) \in \Omega_N (\alpha)$ 
(see Theorem \ref{btt}) and suppress its index. Using
(\ref{eq-transferrel}) and the fact that the norm of a transfer matrix
is equal to the norm of its inverse yields $\|{\cal
  T}^{E_c+\delta}(k,0)\|^{-1} \leq R^{0,E_c+\delta}(k) \le \|{\cal
  T}^{E_c+\delta}(k,0)\|$. Theorem~\ref{btt} therefore guarantees
the existence of a constant $C$ such that for $0\leq k\leq N$ and for $-N^{-\alpha -1/2} <\delta < N^{-\alpha -1/2},$

$$
\frac{1}{C}\;\leq\;
R^{0,E_c+\delta}(k)^2
\;\leq\;
C
\mbox{ , }
\qquad
\frac{1}{C}\;\leq\;
|u^{E_c+\delta}(k)|^2
+|u^{E_c+\delta}(k-1)|^2
\;\leq\;
C
\mbox{ . }
$$

\noindent This readily yields (\ref{eq-efspreading}). 
Now by Lemma~\ref{lem-Prueferderiv}, 

$$
\frac{N}{C}\;\leq\;
{\partial_E}\,\theta^{0,E_c+\delta}(N)
\;\leq\;
N\,C
\mbox{ . }
$$

\noindent Upon integration, one deduces

$$
\frac{N}{C}\,|E-E'|\;\leq\;
|\theta^{0,E}(N)-\theta^{0,{E}'}(N)|
\;\leq\;
N\,C\,|E-E'|
\mbox{ , }
$$

\noindent for all $E,E'\in [E_c-N^{-\alpha-1/2},E_c+N^{-\alpha-1/2}]$. The
oscillation theorem discussed in Section \ref{sec-pruefer} now
gives (\ref{eq-evspacing}).
\hfill $\Box$

\section{Lower bound on dynamics}
\label{sec-dynamics}

The deterministic part of the argument presented in this section 
follows \cite{DT}. Let us return to the simplified notation from
Section~\ref{sec-res} and write $\omega$ instead of $(\omega,l)$ since
based on the results of Section~\ref{sec-deviations} the value of
$l$ will not influence the considerations. Let us begin with some
preliminaries and introduce the Green's function

$$
G^{z}_\omega(n)\;=\;
\left\langle n\left|\frac{1}{H_\omega-z}\right|0\right\rangle
\mbox{ . }
$$

\noindent Note that 

\begin{equation}
\label{eq-greensum}
-t_\omega(n+1)\,G^{z}_\omega(n+1)
+
(v_\omega(n)-E_c-z)\,G^{z}_\omega(n)
-t_\omega(n)\,G^{z}_\omega(n-1)
\;=\;
\delta_{n,0}
\mbox{ . }
\end{equation}

\noindent Using transfer matrices, one now has for $n\leq 0$, 

\begin{equation}
\label{eq-greentransfer1}
\left(
\begin{array}{c}
t_\omega(n)\,G^{z}_\omega(n)
\\
G^{z}_\omega(n-1)
\end{array}
\right)
\;=\;
{\cal T}^z_\omega(n,0)
\left(
\begin{array}{c}
t_\omega(0)\,G^{z}_\omega(0)
\\
G^{z}_\omega(-1)
\end{array}
\right)
\mbox{ , }
\end{equation}

\noindent while for $n\geq 1$, 

\begin{equation}
\label{eq-greentransfer2}
\left(
\begin{array}{c}
t_\omega(n)\,G^{z}_\omega(n)
\\
G^{z}_\omega(n-1)
\end{array}
\right)
\;=\;
{\cal T}^z_\omega(n,1)
\left(
\begin{array}{c}
t_\omega(1)\,G^{z}_\omega(1)
\\
G^{z}_\omega(0)
\end{array}
\right)
\mbox{ . }
\end{equation}

\noindent The following identity is well-known:

\begin{equation}
\label{eq-greentransport}
M_{\omega,q}(T)
\;=\;
\frac{1}{\pi}\,\frac{1}{T}\,
\sum_{n\in\ZZ}
|n|^q\;\int_\RR dE\;|G^{z}_\omega(n)|^2
\mbox{ , }
\qquad
z\;=\;E+\frac{\imath}{T}
\mbox{ . }
\end{equation}

\vspace{.2cm}

\noindent {\bf Proof} of Theorem \ref{theo-lower}.
For given $\alpha>0$ let $c, c'>0$ and $C<\infty$ be the constants
form Theorem~\ref{btt} and choose $N=[c'\,T]$ and $\epsilon =
N^{-1/2-\alpha}$. By Theorem~\ref{btt} there exists
$\Omega_N(\alpha) \subset \Omega$ with $\PP(\Omega_N(\alpha)) =
\Oo({e}^{-cN^{\alpha}})$ and such that for $\omega \in
\Omega_N(\alpha)^c$ one has $\|{\cal
  T}_{\omega}^{E_c+\delta+\imath/T}(n,1)\| \leq C$ for all $|\delta|\leq
N^{-\alpha-1/2}$ and $n\le N$.

For such $\omega,$ because of the uniform bounds on the
matrix elements $t_\omega(n)$ and $v_\omega(n)$, 
for $n=0$ one of the three terms on the l.h.s. of 
(\ref{eq-greensum}) has to be large. Suppose first that
$|G^z_{\omega}(0)|^2 + |G^z_{\omega}(1)|^2 \ge C_6 > 0$, then it
follows from (\ref{eq-greentransfer2}) and
$\|({\cal T}^z_\omega(n,1))^{-1}\|=\|{\cal T}^z_\omega(n,1)\|$
that

$$
\max\left\{|G_\omega^z(n)|^2, |G_\omega^z(n-1)|^2\right\}
\;\geq\;
\frac{C_7}{\|{\cal T}^z_\omega(n,1)\|^2}
\mbox{ . }
$$

\noindent According to the above, as long as $\delta \in [-\epsilon,
\epsilon]$ the transfer matrices
are bounded 
from above by $C$ as long as $n\leq [c_1\,T]$, in which case at least
every second $|G_\omega^z(n)|^2$ is bigger than $C_7/C^2$.
Replacing this into (\ref{eq-greentransport}),

$$
M_{\omega,q}(T)
\,\geq \,
\frac{1}{2\pi T}\,\sum_{0\leq n \leq [c_1\,T]}
n^q\;\int_{[-\epsilon,\epsilon]} 
\!\!\!\!\!d\delta\;\frac{C_7}{C^2}
\, \geq \,
C_8\,T^{q}\epsilon
\,=\,
C_8\,T^{q-\frac{1}{2}-\alpha}
\mbox{ ,}
$$

\noindent for some constant $C_8>0$. If, on the other hand
$|G_{\omega}^z(-1)|^2 \ge C_6 > 0$, then one gets this estimate in the
same way, but based on (\ref{eq-greentransfer1}) instead of
(\ref{eq-greentransfer2}). This uses the fact that the analysis of
Section~\ref{sec-deviations} can also be carried out for ${\cal
 T}^{E_c+\delta}_{\omega}(n,0)$  with negative $n$. A Borel-Cantelli lemma
shows that a.s.\ ${\beta}^-_{\omega,q} \ge
1-(\frac{1}{2}-\alpha)/q$. Since $\alpha >0$ is arbitrary, this
finishes the proof.
\hfill $\Box$

\vspace{.2cm}

\noindent {\bf Proof} of Theorem \ref{theo-lowerdeter}. Follow the
above argument by using the deterministic Proposition
\ref{prop-Lyaptrivial}. 
\hfill $\Box$

\vspace{.2cm}



\begin{thebibliography}{99}

\bibitem[BT]{BT} J.-M. Barbaroux, S. Tcheremchantsev, {\sl Universal
lower bounds for quantum diffusion}, J. Funct. Anal. {\bf 168},
327-354 (1999).

\bibitem[BG]{BG} S. de Bi{\`e}vre, F. Germinet, {\sl Dynamical Localization
for the Random Dimer Schr{\"o}dinger Operator}, J. Stat. Phys. {\bf 98},
1135-1148 (2000).

\bibitem[BL]{BL} P. Bougerol, J. Lacroix, {\sl Products of Random
Matrices with Applications to Schr{\"o}dinger Operators}, (Birkh{\"a}user,
Boston, 1985).

\bibitem[Bov]{Bov} A. Bovier, {\sl Perturbation theory for the random
dimer model}, J. Phys. A: Math. Gen. {\bf 25}, 1021-1029 (1992).

\bibitem[BJ]{bj}J. Bourgain, S. Jitomirskaya, {\sl Anderson localization 
for the band model}, in: Geometric aspects of
functional analysis, Lecture Notes in Math. {\bf 1745}, 67-79, Springer, 
Berlin, 2000.


\bibitem[CK]{CK}
M. Campanino, A. Klein, 
{\sl Anomalies in the one-dimensional Anderson model at weak
disorder}, Commun. Math. Phys. {\bf 130}, 441-456 (1990).

\bibitem[CKM]{CKM} R. Carmona, A. Klein, F. Martinelli,
{\sl Anderson localization for Bernoulli and other singular
potentials}, Comm. Math. Phys.
{\bf 108}, 41-66 (1987).


\bibitem[CS]{CS} V. Chulaevsky, T. Spencer, {\sl Positive Lyapunov
Exponents for Deterministic Potentials}, Commun. Math. Phys. {\bf
168}, 455-466 (1995).


\bibitem[DSS1]{DSS1} D. Damanik, R. Sims, G. Stolz,
{\sl Localization of one dimensional, continuum, Bernoulli-Anderson
models}, to appear in Duke Math. J..

\bibitem[DSS2]{DSS2} D. Damanik, R. Sims, G. Stolz, in preparation.

\bibitem[DT]{DT} D. Damanik, S. Tcheremchantsev, in preparation.

\bibitem[RJLS]{RJLS} R. del Rio, S. Jitomirskaya, Y. Last, B. Simon,
{\sl Operators with singular continuous spectrum: IV. Hausdorff dimension,
rank-one perturbations and localization},
J. d'Analyse Math. {\bf 69}, 153-200 (1996).

\bibitem[DWP]{DWP} D. H. Dunlap, H.-L. Wu, P. W. Phillips,
{\sl Absence of Localization in  Random-Dimer Model},
Phys. Rev. Lett. {\bf 65}, 88-91 (1990).

\bibitem[Fur]{Fur} H. Furstenberg, {\sl Noncommuting random products}, 
Trans. Amer. Math. Soc. {\bf 108}, 377-428 (1963).

\bibitem[GK]{GK} F. Germinet, A. Klein, {\sl Bootstrap Multiscale
Analysis and Localization in Random Media}, Commun. Math. Phys. {\bf
222} 415-448  (2001).

\bibitem[Gua]{Gua} I. Guarneri,
{\sl Spectral properties of quantum diffusion
on discrete lattices}, Europhys. Lett., {\bf 10}, 95-100, (1989); {\sl On an
estimate concerning quantum diffusion in the presence of a fractal spectrum}
, Europhys. Lett., {\bf 21}, 729-733, (1993).


\bibitem[GSB]{GSB} I. Guarneri, H. Schulz-Baldes,
{\sl Intermittent lower bound on quantum diffusion},
Lett. Math. Phys. {\bf 49}, 317-324 (1999).

\bibitem[JSS]{JSS} S. Jitomirskaya, H. Schulz-Baldes, G. Stolz,
Delocalization in polymer models. Preprint. {\tt mp-arc 02-1}.


\bibitem[KS]{KS}
V. Kostrykin, R. Schrader, {\sl Global bounds for the Lyapunov
exponent and the integrated density of states of random
Schr{\"o}dinger operators in one dimension}, J. Phys. A {\bf 33},
8231--8240 (2000).

\bibitem[PTB]{PTB} A. Parisini, L. Tarricone, V. Bellani, G.
Parravicini, E. Diez, F Dominguez-Adame, R. Hey, {\sl Electronic
structure and vertical transport in random dimer}
GaAs-Al$_x$Ga$_{1-x}$As {\sl superlattices}, Phys. Rev. {\bf B 63},
1653218 (2001).

\bibitem[PF]{PF} L. Pastur, A. Figotin, {\sl Spectra of Random and
Almost-Periodic Operators}, (Springer, Berlin, 1992).

\bibitem[PW]{PW} P. Phillips, H.-L. Wu, {\sl Localization and Its
Absence: A New Metallic State for Conducting Polymers}, Science {\bf
252}, 1805-1812 (1992).

\bibitem[SB]{SB}  H. Schulz-Baldes, J. Bellissard, {\sl Anomalous
transport: a mathematical framework}, Rev. Math. Phys. {\bf 10}, 1-46 (1998).

\bibitem[Sim]{Sim} B. Simon, {\sl Absence of ballistic motion},
Comm. Math. Phys. {\bf 134}, 209-212 (1990).

\bibitem[Sim2]{s} B. Simon, {\sl Bounded eigenfunctions and absolutely continuous spectra for
one-dimensional Schroedinger operators}, Proc. Amer. Math. Soc. {\bf 124}, 3361--3369 (1996)


\bibitem[SVW]{SVW} C. Shubin, R. Vakilian, T. Wolff,
{\sl Some harmonic analysis questions suggested by Anderson-Bernoulli
models},  Geom. Funct. Anal.
{ \bf 8} 932-964 (1998).

\end{thebibliography}
\end{document}